\documentclass[iop,onecolumn]{emulateapj}
\usepackage{amssymb,amsmath,amsthm}

\begin{document}

\title{Perspectives on Intracluster Enrichment and the Stellar Initial
  Mass Function in Elliptical Galaxies}
\author{Michael Loewenstein\altaffilmark{1}} \affil{Department of
  Astronomy, University of Maryland, College Park, MD 20742}
\email{Michael.Loewenstein.1@nasa.gov}
\altaffiltext{1}{CRESST and X-ray Astrophysics Laboratory, NASA/GSFC,
Greenbelt, MD.}

\begin{abstract}

Stars formed in galaxy cluster potential wells must be responsible for
the high level of enrichment measured in the intracluster medium
(ICM); however, there is increasing tension between this truism and
the parsimonious assumption that the stars in the generally old
population studied optically in cluster galaxies emerged from the same
formation sites at the same epochs. We construct a phenomenological
cluster enrichment model to demonstrate that ICM elemental abundances
are underestimated by a factor $>2$ for standard assumptions about the
stellar population -- a discrepancy we term the ``cluster elemental
abundance paradox''. Recent evidence of an elliptical galaxy IMF
skewed to low masses deepens the paradox. We quantify the adjustments
to the star formation efficiency and initial mass function (IMF), and
SNIa production efficiency, required to resolve this while being
consistent with the observed ICM abundance pattern. The necessary
enhancement in metal enrichment may, in principle, originate in the
observed stellar population if a larger fraction of stars in the
supernova-progenitor mass range form from an initial mass function
(IMF) that is either bottom-light or top-heavy, with the latter in
some conflict with observed ICM abundance ratios. Other alternatives
that imply more modest revisions to the IMF, mass return and remnant
fractions, and primordial fraction, posit an increase in the fraction
of $3-8~{\mathrm M}_{\sun}$ stars that explode as SNIa or assume that
there are more stars than conventionally thought -- although the
latter implies a high star formation efficiency.  We discuss the
feasibility of these various solutions and the implications for the
diversity of star formation in the universe, the process of elliptical
galaxy formation, and the origin of this ``hidden'' source of ICM
metal enrichment.

\end{abstract}

\section{Context}

The hot plasma that pervades the volume of the most massive galaxy
clusters -- the intracluster medium (ICM) -- provides a wealth of
diagnostic data on the process of galaxy formation in structures
formed from the largest primordial density fluctuations to have
entered the nonlinear regime and undergone gravitational collapse. The
history and efficiency of star formation, and the effects of
interactions among galaxies and between galaxies and the environment
in the form of infall, outflow, and dynamical stripping, are reflected
in the thermal and chemical properties of the ICM -- both in
individual systems and in the evolving population of clusters.

In massive ($>10^{14}~{\mathrm M}_{\sun}$) clusters, the ICM dominates
the baryon inventory and accounts for $>10$\% of the total matter
content (e.g., Lagan\'a et al. 2011). The discovery that cluster gas
fractions expected to be representative of the universe as a whole
exceed $\Omega_{baryon}$\footnote{$\Omega$ represents mean densities
  in units of the critical density that delineates closed and open
  universes.} precipitated a crisis referred to as the ``baryon
catastrophe'' when combined with the assumption of a matter-dominated
universe and evidence for, and the inflationary prediction of, a flat
universe -- {\it i.e.} $\Omega_{matter}=1$
\citep{fabian1991,white1993}. Of course this paradox was resolved by
the discovery of dark energy and the concordance cosmology which
reconciles a flat universe with a reduced $\Omega_{matter}$. The
cosmic baryon matter fraction is now accurately determined,
$\Omega_{baryon}/\Omega_{matter}=0.17$ \citep{jarosik2011}, with a
relatively modest percentage of the baryons collapsed into stars,
$\Omega_{stars}/\Omega_{matter}\approx 0.0074-0.011$
\citep{gallazzi2008}.

However a related paradox persists to this day. While a solar
nucleosynthetic yield (i.e., star formation ultimately producing the
amount of metals necessary to enrich one solar mass to solar
abundances) is sufficient only to enrich baryons to average abundance
of $<0.1$ on a universal scale, cluster baryons are enriched in Fe
(and other elements) to $\sim$half-solar
\citep{tamura2004,dePlaa2007,leccardi2008,bregman2010,matsushita2011,baldi2012,andreon2012}.\footnote{We
  adopt the solar abundance standard of \citet{asplund2009} where
  Fe/H$=3.16\times 10^{-5}$ by number.} Star formation in cluster
galaxies is evidently more efficient than in the field; however, a
discrepancy between cluster metals and the number of stars evidently
available to produce these metals was noted following the very first
detection of intraclcuster Fe by \cite{mitchell1976} and
\cite{serlemitsos1977}
\citep{vigroux1977,rothenflug1984,ra1985,arnaud1992}. This discordance
was confirmed and quantified in detail following the groundbreaking
accuracy and range of abundance measurements made with the {\it ASCA}
X-ray Observatory
\citep{loewenstein1996,mushotzky1997,pagel1999,pagel2002,lin2003,finoguenov2003,portinari2004,lin2004,delucia2004,loewenstein2006,maoz2010,bregman2010}. This
may be framed in terms of the Fe-mass-to-light ratio
\citep{arnaud1992,renzini1993}: for $M_{*}/L_B=5$ a solar yield
corresponds to $M_{Fe}/L_B\sim 0.0065$ -- falling short by a factor of
5 or more compared to what is measured \citep{sakuma2011,sato2012}.

To resolve this ``cluster elemental abundance paradox'' one generally
must conclude that there was more star formation in clusters than
conventionally estimated, and/or that star formation in galaxy
clusters has an enhanced efficiency of producing supernova progenitors
and synthesizing metals. ``Missing'' stars may take the form of low
surface brightness intracluster light (ICL), inferred both from
observations \citep{zaritsky2004,lin2004,gonzalez2007,kb2007} and
simulations \citep{puchwein2010,rudick2011}. Recent literature
includes a large range in calculated star-to-ICM ratios primarily due
to divergent ICL estimates, but also to different assumed stellar
mass-to-light ratios. A stellar initial mass function (IMF) that is
relatively top heavy increases both the ratio of stars formed, and of
metals created, to present-day light. Such an IMF may be bimodal in
nature \citep{elbaz1995,larson1998,moretti2003}, the second mode
perhaps associated with a distinct pre-enrichment population
\citep{bregman2010} where Population III hypernovae
\citep{loewenstein2001} may play a role. Given the conventional wisdom
that most ICM metals originate in elliptical galaxies
\citep{arnaud1992}, this problem clearly connects to the fundamental
galaxy formation questions of the IMF in ellipticals and the transport
of material from these galaxies into intergalactic space.

In this paper we undertake a fresh and comprehensive, though generic,
examination of the metal inventory in the ICM of rich galaxy clusters.
We focus on addressing a single well-defined, though multifaceted,
question: {\it what characteristics of the stellar population are {\bf
    necessary} to produce the observed level of ICM enrichment?} In
doing so we address issues related to the IMF, star and galaxy
formation efficiency, galactic winds, the astrophysics of supernova
progenitors and explosions, and the apportionment of products of
different supernova types into stars and ICM. Section 2 quantitatively
summarizes the cluster elemental abundance paradox.

In Section 3, where a standard IMF is assumed, the level of enrichment
and abundance pattern are related to phenomenological parameters that
encapsulate the star formation efficiency, and the demographics of
supernovae and the success of stars in locking up the products of
their explosion. Substantial departures from standard values are
required to match observations. In Section 4, we cast a wider net by
considering the effects on ICM enrichment of a wide range of IMFs in
the context of a self-consistent galaxy chemical evolution treatment
that accounts for the relevant astrophysical quantities. When
juxtaposed with recent evidence for an IMF in elliptical galaxies that
produces {\it fewer} metals than a local IMF, this analysis reinforces
and clarifies the conflict between the ICM metallicity and the
characteristics of the stellar population generally assumed
responsible for ICM enrichment. Results and their implications are
discussed, and conclusions summarized, in Section 5.

\section{The Cluster Elemental Abundance Paradox Quantified}

\subsection{Basics}

Consider the total baryon mass in a cluster of galaxies within some
sufficiently large radius that it may be considered a closed box in
the chemical evolution sense -- that is, all products resulting from
the transformation of some of this gas into stars (including the stars
themselves) are contained within this radius. There is direct evidence
that this is a good approximation for sufficiently massive clusters if
invoked at radii that are a significant fraction of the virial radius,
based on the consistency between the total cluster baryon fraction and
the universal value mentioned above \citep{landry2012}, {\it perhaps}
with a $\sim 10$\% ``depletion'' correction at $r_{500}$ (the radius
within which the average mass density is 500 times the critical
density); see
\cite{gonzalez2007,pratt2009,giodini2009,planck2013,planelles2013,eckert2013}. Presumably
this is a result of the extreme depth of their gravitational potential
wells. However it is possible that this also applies to galaxy groups,
and perhaps even giant elliptical galaxies if one could inventory the
gas all the way out to the virial radius (and perhaps beyond, if
entropy injection has dispersed the gas distribution).

We define the overall efficiency of converting gas into stars,
$\varepsilon_{sf}$, such that the total mass in stars formed
(regardless of where) is
\begin{equation}
M_{*,form}=\varepsilon_{sf}\, M_{baryon},
\end{equation}
where $M_{baryon}$ is the total baryon mass being considered. At
the present time the total mass in stars, whether contained in
individual cluster galaxies (including the brightest cluster galaxy --
BCG) or associated with intracluster light (ICL), is
\begin{equation} 
M_{*}=M_{baryon}\, \varepsilon_{sf}\, (1-r_{*}),
\end{equation} 
and the mass in gas is
\begin{align} 
M_{gas}& = M_{baryon}-M_{*}\\ & =M_{baryon}\left[1-\varepsilon_{
    sf}(1-r_{*})\right],
\end{align}
where $r_{*}$ is the stellar ``mass return fraction'' -- the fraction
of the mass previously formed into stars recycled back into gas. For
massive clusters one may neglect the distinction between the total
mass in gas and the mass in the ICM, and henceforth we equate
$M_{gas}$ with $M_{ICM}$. The star formation efficiency, in terms of
the observable $M_{ICM}/M_{*}$ is
\begin{equation}
\varepsilon_{
  sf}=(1-r_{*})^{-1}\left(1+\frac{M_{ICM}}{M_{*}}\right)^{-1}.
\end{equation}

We consider the enrichment of cluster baryons in chemical elements
released in supernova explosions, i.e. those of atomic number $A\ge
8$. It is straightforward to extend the analysis to elements
statically synthesized in intermediate mass stars; however, the
elements (C, N) to which this applies are not well-constrained by
current X-ray observations.

Both the overall level of baryon enrichment (that is, the metallicity)
and the abundance pattern are determined by the total number of
supernovae and their nucleosynthetic yields. We separate the
enrichment contributions of the two main classes of supernovae -- Type
Ia supernovae (SNIa) that result from the explosion of a white dwarf,
and core collapse supernovae (SNcc). Their total numbers may be
expressed as
\begin{equation}
N^{cc}=\eta^{cc}\, M_{*,form}
\end{equation}
and
\begin{equation}
N^{Ia}=\eta^{Ia}M_{*,form};
\end{equation}
\noindent where $\eta^{cc}$ and $\eta^{Ia}$ are, respectively, the
specific numbers of SNcc and SNIa explosions per star formed. It is
useful to define the total supernova number, supernova ratio, and SNIa
fraction as follows:
\begin{align} 
N^{SN}& =N^{cc}+N^{Ia}\\& =\eta^{SN}~(1-r_{*})^{-1}M_{*},
\end{align}
where
\begin{equation}
\eta^{SN}=\eta^{cc}+\eta^{Ia};
\end{equation}
\begin{equation}
R^{SN}\equiv\frac{\eta^{Ia}}{\eta^{cc}};
\end{equation}
and 
\begin{equation}
f^{Ia}\equiv\frac{\eta^{Ia}}{\eta^{SN}}=\frac{R^{SN}}{1+R^{SN}}.
\end{equation}
Note that the number of SN per unit mass in the ICM is
\begin{equation}
\frac{N^{SN}}{M_{ICM}}=\eta^{SN}~(1-r_{*})^{-1}\frac{f_{*}}{f_{ICM}},
\end{equation}
where $f_{*}$ and $f_{ICM}$ are the present-day mass fractions
of stars and gas: $f_{*}/f_{ICM}=M_{*}/M_{ICM}$. 

\subsection{Stars and Supernovae}

A combination of theoretical and empirical considerations enter into
determination of the stellar and supernovae parameters. Star formation
is not sufficiently well-understood to allow an {\it a priori}
estimate of $\varepsilon_{sf}$, and $M_*$ is estimated from
observations of cluster starlight in galaxies and intracluster
space. To convert to mass, stellar population synthesis can be
employed, involving assumptions about the IMF -- among many other
factors. The mass-to-light ratio in individual galaxies can also be
inferred from dynamical modeling of stellar velocity dispersion
distributions. The mass return fraction, $r_{*}$, may be calculated
from the IMF, star formation history (SFH), and relation of stellar
remnant mass to progenitor mass for individual stars estimated from
stellar evolution theory and observations in the Galaxy. The SNcc
efficiency, ${\eta^{cc}}$, depends on the IMF and range of masses that
result in core collapse explosions. Although one may model the
time-dependence of the SNIa rate from assumptions about the binary
star progenitor configuration and the distributions of binary mass
ratios and separations, the normalization is difficult to estimate
{\it a priori}. Empirical estimates of ${\eta^{Ia}}$ (and
${\eta^{cc}}$) are emerging from supernova surveys that also constrain
the distribution of delay times from the observed evolution of the
SNIa rate \citep{maoz2012,sand2012}. Estimation of these fundamental
quantities -- $M_*$, $r_{*}$, ${\eta^{Ia}}$, ${\eta^{cc}}$ -- in
principle requires a convolution and a time-integration over an
evolving galaxy population with disparate SFHs and, conceivably,
IMFs. A first order approximation treats the ensemble stellar
population as a single simple population of stars formed at some
(high) redshift with a common IMF -- an approximation most suitable
for rich clusters where the total galaxy mass is most dominated by
elliptical galaxies.

We can insert some reasonable values to get a sense for the expected
level of supernova enrichment and relative contribution from the two
classes of supernovae. For a ``diet Salpeter'' IMF that represents a
simple alteration -- proposed as a means of reconciliation with the
observed relative frequency of $\sim$subsolar mass stars
\citep{bell2001} -- of the classic single-slope Salpeter function
\citep{salpeter1955}, $\eta^{Ia}\sim 0.002$ \citep{maoz2012},
$r_{*}\sim 0.35$ for an old stellar population \citep{fardal2007,
  O'Rourke2011}, and $\eta^{cc}\sim 0.008$
\citep{maoz2012,botticella2012,dahlen2012}.

One then predicts $R^{SN}\sim 0.25$ ($f^{Ia}\sim 0.2$) and
$\eta^{SN}\sim 0.01$. For massive clusters (i.e., $M_{500}\equiv
M(r_{500})>10^{14}~{\mathrm M}_{\sun}$), recent studies report a range
in stellar mass fraction evaluated at $r_{500}$, reflecting different
treatments of ICL and in conversion from light to mass
\citep{zhang2011}, with $f_{*}/f_{ICM}$ typically $\sim 0.1$
\citep{lin2003,gonzalez2007,giodini2009,ettori2009,dai2010,andreon2010,bregman2010,lagana2011,balogh2011,zhang2011,lin2012}
(though with substantial systematic uncertainty and study-to-study
variation; see \cite{leauthaud2012}), and evidence of an increase in
magnitude and scatter with decreasing cluster mass. The resulting
total number of supernova explosions per solar mass of ICM is $\sim
1.5\times 10^{-3}(10f_{*}/f_{ICM})~{\mathrm M}_{\sun}^{-1}$.
 
\subsection{Application to the ICM}

The equation for the mass of the $ith$ element in the ICM, $M_i$, in
terms of the number of SNIa and SNcc that enrich the ICM, $N^{Ia}$ and
$N^{cc}$, and the yields per SNIa and SNcc, $y_i^{Ia}$ and $\langle
y_i^{cc}\rangle$, is
\begin{equation}
M_i=N^{cc}\langle
y_i^{cc}\rangle+N^{Ia}y_i^{Ia}=N^{SN}(1+R^{SN})^{-1}(\langle
 y_i^{cc}\rangle +R^{SN}y_i^{Ia}),
\end{equation}
where, as defined above, $R^{SN}\equiv N^{Ia}/N^{cc}$ and
$N^{SN}\equiv N^{Ia}+N^{cc}$. The IMF($\phi$)-averaged SNcc yield is 
\begin{equation}
\langle y_i^{cc}\rangle= \frac{\int^{m_{up}}_{m_{cc}}dm\phi(m)
  y_i^{cc}(m)}{\int^{m_{up}}_{m_{cc}}dm\phi(m)},
\end{equation}
where $m_{cc}$ and $m_{up}$ are the lower and upper limits for the
masses of SNcc progenitors, and a single universal set of SNIa yields
is assumed. Despite a plethora of IMF parameterizations
\citep{kroupa2012}, there is general consensus that a Salpeter slope
\citep{salpeter1955}, $\phi\sim m^{-2.35}$ applies at the high mass
end relevant for SNcc -- at least for star formation under ``normal''
conditions.

The resulting mass fraction in the ICM
(mass $M_{ICM}$) of the $ith$ element, $f_i$, is
\begin{equation}
f_i\equiv \frac{M_i}{M_{
    ICM}}=\frac{N^{SN}}{M_{ICM}}(1+R^{SN})^{-1}(\langle
y_i^{cc}\rangle +R^{SN}y_i^{Ia}),
\end{equation}
and the mass fraction relative to the solar mass fraction,
\begin{equation}
\frac{f_i}{f_{i\sun}}=\frac{N^{SN}}{M_{ICM}}(1+R^{SN})^{-1}
\left(\frac{\langle y_i^{cc}\rangle}{f_{i\sun}}
+R^{SN}\frac{y_i^{Ia}}{f_{i\sun}}\right).
\end{equation}
The relationship between mass fraction, $f_i$, and abundance, $z_i$,
of the $ith$ element (the number of atoms of element $i$ relative to
that of H, i.e. the entries in standard abundance tables) is
$z_i=(f_i/X)(A_i/A_H)^{-1}$, where $X$ and $A_H$ are the hydrogen mass
fraction and atomic weight ($A_H=1.008$ AMU) and $A_i$ the atomic
weight of the $ith$ element. Relative to solar, the abundance is
\begin{equation} 
\frac{z_i}{z_{i\sun}}=\frac{X_{\sun}}{X}\frac{f_i}{f_{i\sun}}\approx
\frac{f_i}{f_{i\sun}},
\end{equation}
where $f_{i\sun}=z_{i\sun}X_{\sun}(A_i/A_H)$, and the approximation
$X=X_{\sun}$ is invoked -- an approximation that is valid as long as
the total mass fraction of metals ($<2$\% for solar abundances) is
small and the He abundance is fixed. $X$ for various solar standard
abundance sets is given in Table 4 of \cite{asplund2009}.

Finally, the abundance relative to solar is expressed as 
\begin{equation} 
Z_i\equiv
\frac{z_i}{z_{i\sun}}=\frac{N^{SN}}{M_{ICM}}(1+R^{SN})^{-1}(\langle
y_i^{cc'}\rangle+R^{SN}y_i^{Ia'})
\end{equation}
where ${N^{SN}}/{M_{ICM}}$ is given by equation (13); and,
$y_i^{cc'}\equiv \langle y_i^{cc}\rangle/f_{i\sun}$ and
$y_i^{Ia'}\equiv y_i^{Ia}/f_{i\sun}$. For $M_{ICM}$ measured in
M$_{\sun}$ these are the yields of the $ith$ element relative to the
mass of that element contained in one ${\mathrm M}_{\sun}$ of solar
abundance material. As noted above for the total baryons, $R^{SN}$ and
$N^{SN}$ fully determine the level and pattern of ICM enrichment --
modulo sets of SNIa and SNcc yields - and can be compared to ICM
abundances. The new approach of \citet{bsl2012} directly fits X-ray
spectra to a model parameterized by $R^{SN}$ and $N^{SN}$ via the
abundance predicted by equation (19).

Focusing on Fe, the element with the most widely determined and most
accurate global ICM abundance measurement, equation (19) predicts
$Z_{Fe,ICM}=0.255(10f_{*}/f_{ICM})$, for the values of $R^{SN}$ and
$N^{SN}$ derived at the end of the previous subsection and adopting
($y_i^{Ia},\langle y_i^{cc}\rangle$)=($0.743~{\mathrm
  M}_{\sun},0.0825~{\mathrm M}_{\sun}$) from \citet{kobayashi2006} --
about half the typical observed value for $f_{*}/f_{ICM}=0.1$. However
this assumes that all of the metals produced by supernovae reside in
the ICM. The values $R^{SN}$ and $N^{SN}$ relevant here correspond to
those supernova explosions that enrich the ICM (or, for some
particular X-ray measurement, those in a particular spectral
extraction region of a particular cluster). Not all of the products
resulting from supernova nucleosynthesis are available to enrich the
ICM.

\subsection{Metals Locked Up in Stars}

One approach to evaluating galaxy cluster enrichment is to estimate
the total inventory of metals in stars and in the ICM in the context
of the total required number of supernova explosions. However, our
focus here will be on the ICM which offers more accurate abundance
determinations over a wider range of elements via X-ray
spectroscopy. This specifically requires a correction accounting for
how supernova products are apportioned among gas and stars.

The galactic mass in rich clusters is dominated by early-type systems
that form their stars rapidly. This results in the well-established
enhancement in $[\alpha/Fe]$, the abundance ratio of $\alpha$-elements
to Fe (expressed as the logarithm with respect to solar) -- i.e., SNcc
products are preferentially locked up in stars. In their investigation
of the giant elliptical galaxy NGC 4472, \citet{loewenstein2010} found
that a ratio of SNIa to total supernovae of $N^{Ia}_*/N^{SN}_*\sim
0.11$ ($N^{cc}_*/N^{SN}_*\sim 0.89$) and number of supernova per mass
in (present-day) stars of $N^{SN}_*/M_{*}\sim 0.0083$ resulted in
$[\alpha/Fe]_*\sim 0.25$ and $Z_{Fe,*}\sim 1$ (as observed in this
particular galaxy, but typical of the class; see also Lin et al. 2003,
Gallazzi et al. 2008). This enables us to estimate the lock-up
corrections, $\eta^{cc}_*$ and $\eta^{Ia}_*$, needed to convert the
specific supernova numbers to those available for enrichment of the
ICM as follows:
\begin{equation}
\eta^{Ia}_*=\frac{N^{Ia}_*}{M_*}(1-r_{*})\approx 6.0\times
10^{-4}Z_{Fe,*},
\end{equation}
and
\begin{equation}
\eta^{cc}_*=\frac{N^{cc}_*}{M_{*}}(1-r_{*})\approx 4.8\times
10^{-3}Z_{Fe,*}.
\end{equation}
The corresponding values available to enrich the ICM,
$\eta^{Ia}_{ICM}$ and $\eta^{cc}_{ICM}$, are
\begin{equation}
\eta^{Ia}_{ICM}=\eta^{Ia}-\eta^{Ia}_*=1.4\times 10^{-3}~{\mathrm
  M}_{\sun}^{-1},
\end{equation}
and
\begin{equation}
\eta^{cc}_{ICM}=\eta^{cc}-\eta^{cc}_*=3.2\times 10^{-3}~{\mathrm M}_{\sun}^{-1}
\end{equation}
for the default parameters considered above. That is, the metal
production from $\sim 60$\% of SNcc and $\sim 30$\% of SNIa must be
locked up in stars to enrich them to solar Fe abundances and
$[\alpha/Fe]_*\sim 0.25$ -- with these factors deducted from the total
to obtain the effective enrichment of the ICM. {\it A relative
  overabundance of SNIa contributing to ICM enrichment is expected
  based on the inference that the accelerated formation of stars in
  clusters preferentially locks up the products of SNcc.} This
asymmetry is observed in the abundance patterns in cluster cores
\citep{dePlaa2007,lovisari2011}, although whether this extends
globally remains an open question -- e.g., the smothering of galactic
winds in central dominant galaxies may skew the pattern, relative to
the ICM as a whole, via concentrated direct injection of SNIa.

Based on these estimates, for the supernovae remaining available to
enrich the ICM, $R^{SN}\sim 0.44$ ($f^{Ia}\sim 0.30$) and $N^{SN}/M_{
  ICM}\sim 0.71(10f_{*}/f_{ICM})\times 10^{-3}{~{\mathrm
    M}_{\sun}}^{-1}$. This enriches the ICM in Fe to the level
$Z_{Fe,ICM}=0.155(10f_{*}/f_{ICM})$, thus quantifying the paradox
that baryons in clusters of galaxies are enriched beyond what is
expected based on the stars we see in galaxies today -- unless either
the star formation efficiency exceeds that in the field by a factor of
$\sim 3$, or supernovae are more efficiently produced per unit star
formation.

\section{A More Comprehensive Examination (I)}

In this, and subsequent, sections we investigate stellar and ICM
abundance predictions for a range of elements -- focusing on a subset
selected on the basis of a combination of accessibility and diagnostic
power: O, Mg, Si, Fe, and Ni. First, we consider a wide range of
published yield sets and apportionment of metals into stars and ICM in
an effort to place robust constraints on the required efficiency of
star formation. In order to be as general and assumption-free as
possible we introduce two parameters that gauge the efficiency with
which stars may lock up supernova products and do so {\it
  asymmetrically}, i.e. preferentially for SNcc relative to SNIa. We
consider the specific effects of varying the IMF, with its coupled
impact on $r_{*}$, $\eta^{cc}$, and $\eta^{Ia}$, in the following
section, fixing these parameters at the values described above
($r_{*}\sim 0.35$, $\eta^{Ia}\sim 0.002$, $\eta^{cc}\sim 0.008$) for
immediate purposes. Note that with these parameters, supernovae enrich
cluster baryons in Fe to a mass-averaged abundance of
$Z_{bar,Fe}=1.66\varepsilon_{sf}$.

We generalize our treatment of quantifying the fraction of metals
synthesized by supernovae that are inaccessible for ICM enrichment due
to lock-up in stars by defining a SNcc lock-up fraction, $\beta^{cc}$,
and supernova asymmetry parameter, $\alpha^{SN}$, such that
\begin{equation}
\beta^{cc}\equiv \frac{\eta^{cc}_*}{\eta^{cc}},
\end{equation}
and
\begin{equation}
\alpha^{SN}\equiv \left(\frac{R^{SN}_*}{R^{SN}}\right)^{-1},
\end{equation}
where $R^{SN}_*\equiv\eta^{Ia}_*/\eta^{cc}_*$. By definition,
$\beta^{cc}\le 1$, and $\alpha^{SN}\ge 1$ is expected under conditions
where rapid conversion of gas into stars results in preferential
incorporation into stars of SNcc products with respect to those from
SNIa products that are released over a relatively extended time
interval. We provisionally adopt the values that correspond to the
estimates of the previous section as standard for the remainder of
this section: $\beta^{cc}=0.6$ and $\alpha^{SN}=2$. ICM abundances are
calculated, for a given star formation efficiency $\varepsilon_{sf}$,
from ICM-specific versions of equations (5), (10), (11), (13), (19),
with the supernovae per star formed effectively reduced to
\begin{equation}
\eta^{Ia}_{ICM}=\eta^{Ia}\left(1-\frac{\beta^{cc}}{\alpha^{SN}}\right),
\end{equation}
and
\begin{equation}
\eta^{cc}_{ICM}=\eta^{cc}(1-\beta^{cc}).
\end{equation}
The results, assuming \citet{kobayashi2006} supernova nucleosynthetic
yields,\footnote{Although these are averages for an IMF with a
  Salpeter slope at the high mass end, we adopt these in general --
  the differences for the IMFs we consider are generally small.} are
displayed in Figure 1. Figures 1a and 1b confirm that the default (IMF
and) lock-up parameters predict a stellar population enriched to solar
Fe abundances, and ICM Fe abundances lower than observed for
$\varepsilon_{sf}\sim 0.15$ ($f_{*}/f_{ICM}=0.11$). One may recover
the observed level of ICM Fe enrichment for $\varepsilon_{sf}\sim 0.3$
($f_{*}/f_{ICM}=0.24$), but only for extreme values of the lock-up
parameters that imply that most of the Fe produced by stars ends up in
the ICM, and that stellar Fe abundances are well below solar -- in
contradiction to observations.

\begin{figure}[h]
\includegraphics[scale=0.25,angle=0]{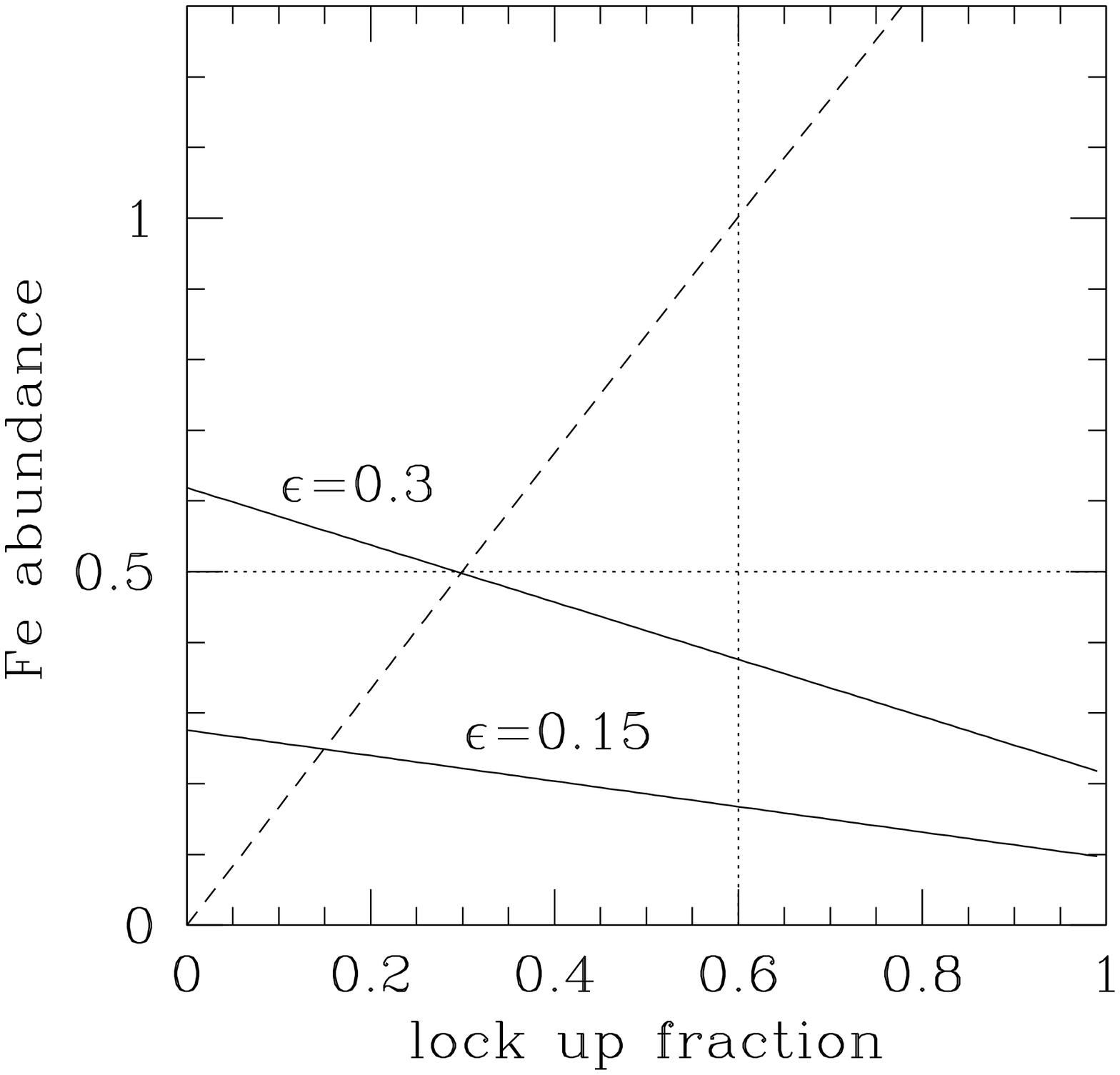}
\hfil
\includegraphics[scale=0.25,angle=0]{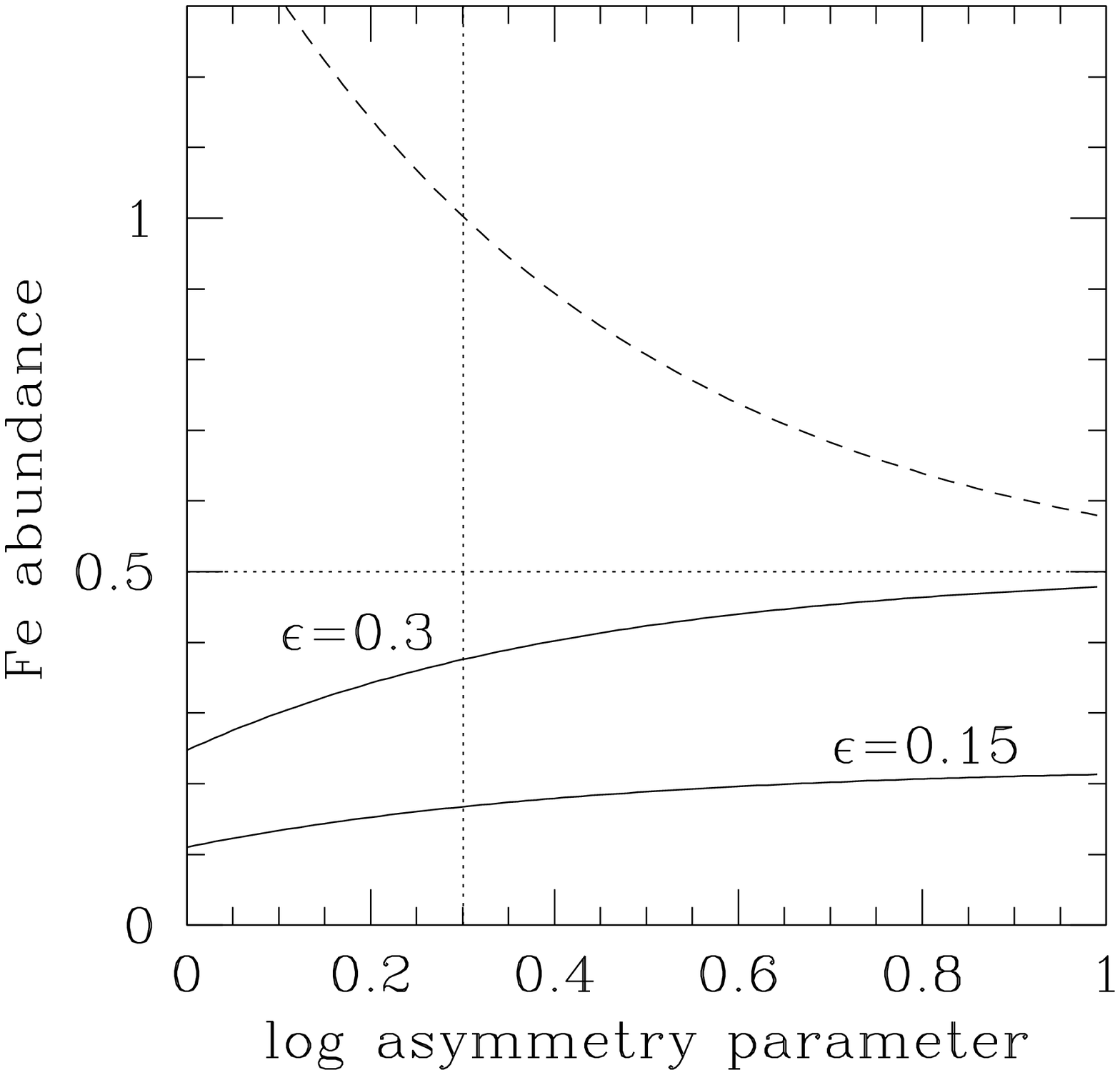}
\hfil
\includegraphics[scale=0.25,angle=0]{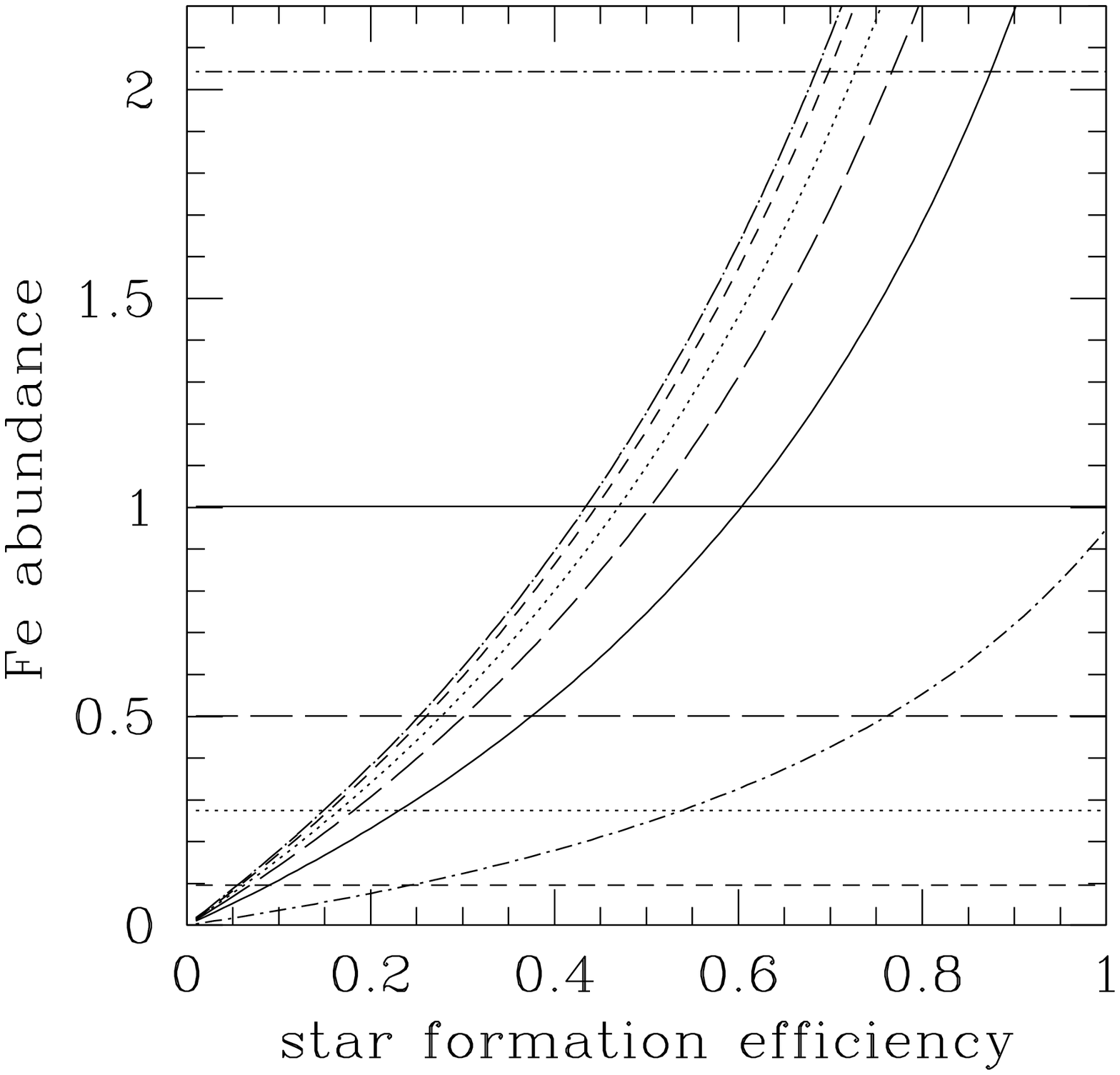}
\hfil
\caption{\footnotesize{{\it Left} panel (a): ICM (solid lines) and
    stellar (broken line) Fe abundance versus lock-up fraction
    $\beta^{cc}$ for (the adopted standard) $\alpha^{SN}=2$, and
    $\varepsilon_{sf}=0.15$ or 0.30. The horizontal dotted line shows
    the typical observed value abundance, $Z_{Fe,ICM}=0.5$, the
    vertical dotted line the adopted standard $\beta^{cc}=0.6$. {\it
      Middle} panel (b): Same as (a) versus $\alpha^{SN}$ for (the
    adopted standard) $\beta^{cc}=0.6$. The vertical dotted line shows
    the adopted standard $\alpha^{SN}=2$. {\it Right} panel (c): ICM
    (curves) and stellar (horizontal lines) Fe abundance versus
    $\varepsilon_{sf}$ for pairs
    ($\alpha^{SN},\beta^{cc}$)=(10,0.1)(short-dashed),
    (3,0.2)(dotted), (2,0.6)(standard: solid), (2.0,0.3)(long-dashed),
    and (1,0.8)(dot-short-dashed). The dot-long-dashed line shows the
    limiting case $\beta^{cc}\rightarrow 0$. }}
\end{figure}

Figure 1c plots $Z_{Fe,ICM}$ and $Z_{Fe,*}$ versus $\varepsilon_{sf}$
for selected pairs ($\beta^{cc}$, $\alpha^{SN}$) -- including the
default. Also
shown is the limiting case $\beta^{cc}\rightarrow 0$ where 100\% of
metals produced by stars reside in the ICM. In this case
\begin{equation}
Z_{Fe,ICM}=\frac{1.66\varepsilon_{sf}}{1-\varepsilon_{sf}(1-r_{*})},
\end{equation}
from which one can see that $\varepsilon_{sf}>0.25$ represents an
absolute lower limit to the star formation efficiency required to
enrich the ICM to $Z_{Fe,ICM}>0.5$. This figure provides an
alternative demonstration that, for $\varepsilon_{sf}=0.15$,
$Z_{Fe,ICM}<0.3$ even for extreme models where such a large fraction
of supernova-produced metals is released into the ICM that
insufficient metals remain available to enrich the stars to the
observed level. {\it Both} relatively large star formation, and small
lock-up efficiencies\footnote{or, more precisely for Fe,
  $\beta^{cc}/\alpha^{SN}<<1$} are required to simultaneously enrich
the stars and ICM to the observed level (see, also, Sivanandam et
al. 2009).

The increasing divergence of stellar and ICM abundance ratios (that
are independent of $\varepsilon_{sf}$) with increasing $\alpha^{SN}$
is shown for $\beta^{cc}=0.6$ in Figure 2a. One can see how, in this
case, the enhanced $[\alpha/Fe]$ measured in the old stellar
populations that dominate cluster galaxies implies a large asymmetry
parameter and, as a result, subsolar ratios of $\alpha$-elements with
respect to Fe in the ICM. Since $[\alpha/Fe]_{ICM}<<1$ is not
observed, large values of $\alpha^{SN}$ may be ruled out. This
divergence narrows with decreasing $\beta^{cc}$ and $\sim$solar ICM
abundance ratios emerge for $\beta^{cc}\sim 0.3$, i.e a lower lock-up
fraction (Figure 2b).

\begin{figure}[h]
\includegraphics[scale=0.33,angle=0]{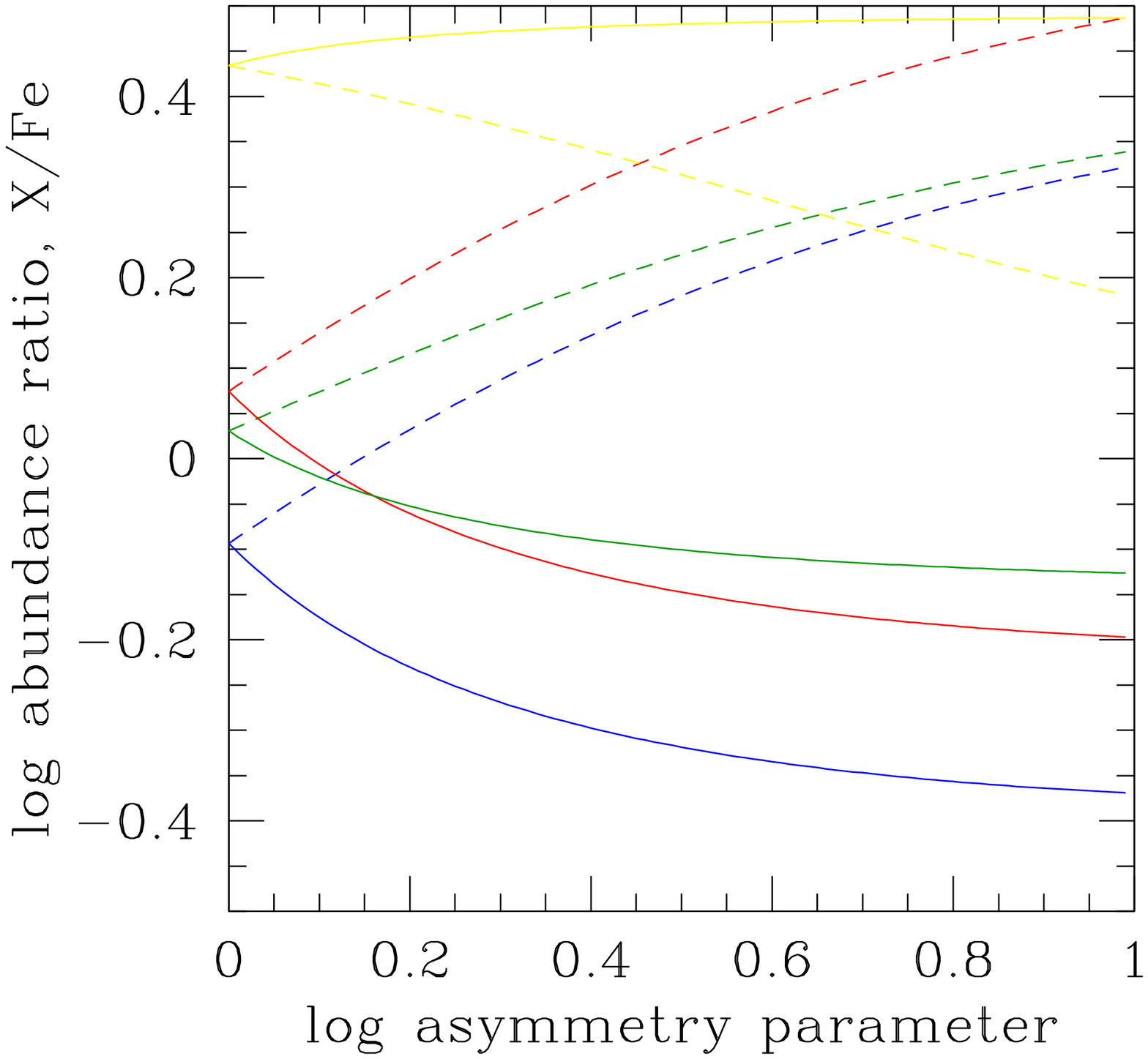}
\hfil
\includegraphics[scale=0.33,angle=0]{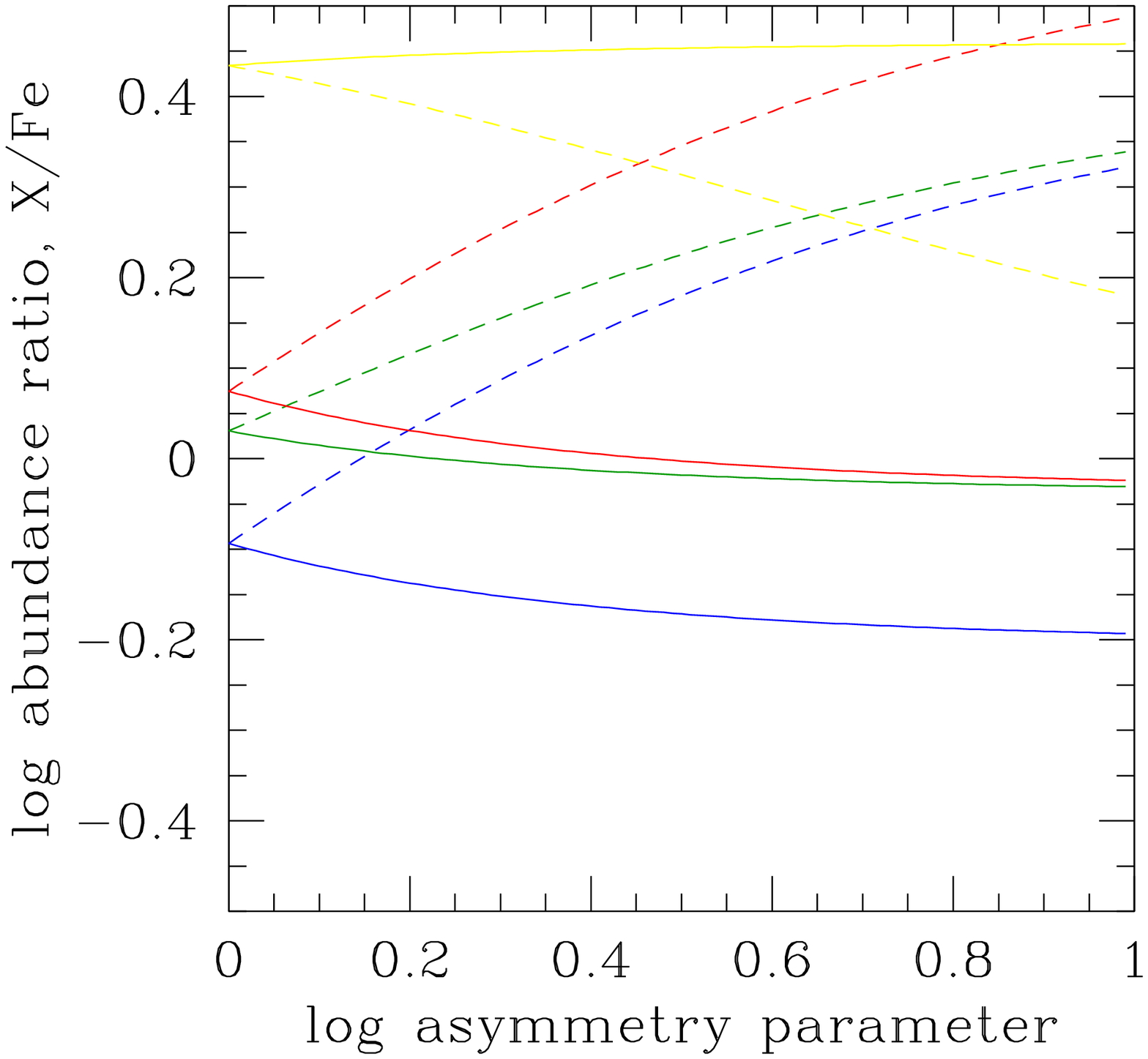}
\caption{\footnotesize{{\it Left} panel (a): ICM (solid lines) and
    stellar (broken lines) abundance ratios with respect to Fe for O
    (red), Mg (blue), Si (green), and Ni (yellow) versus $\alpha^{SN}$
    for the standard $\beta^{cc}=0.6$. {\it Right} panel (b): same
    as (a) for $\beta^{cc}=0.3$.}}
\end{figure}

The effects on cluster enrichment of adopting different SNIa yield
sets is shown in Figures 3-5. The solid lines correspond to previous
plots that utilize \citet{kobayashi2006} yields, the dotted lines use
the same SNcc yields, but alternative SNIa yield sets from
\cite{nom97,mae2010} -- see \cite{loewenstein2010}. Here we focus on
Fe, Ni, and Si (O and Mg are always dominated by SNcc enrichment and
insensitive to this choice). Figure 3 shows the total baryon
enrichment, which is independent of $\beta^{cc}$ and $\alpha^{SN}$,
and demonstrates that star formation efficiencies
$\varepsilon_{sf}\sim 0.25-0.5$ are required to enrich cluster baryons
to a relatively modest Fe abundance of half-solar --
$\varepsilon_{sf}\sim 0.4-0.7$ to attain 0.75 solar. Figures 4 and 5
that, respectively, show the ICM abundances for the standard lock-up
parameters, and the maximum ICM abundances corresponding to zero
stellar metallicity, demonstrate that the conclusions about the
required efficiency of star formation are not a result of a particular
choice of yields sets. Using SNcc yields from \cite{woosley1995}
(their standard explosion energy, solar abundance progenitor model)
does not alter these conclusions.

\begin{figure}[h]
\includegraphics[scale=0.25,angle=0]{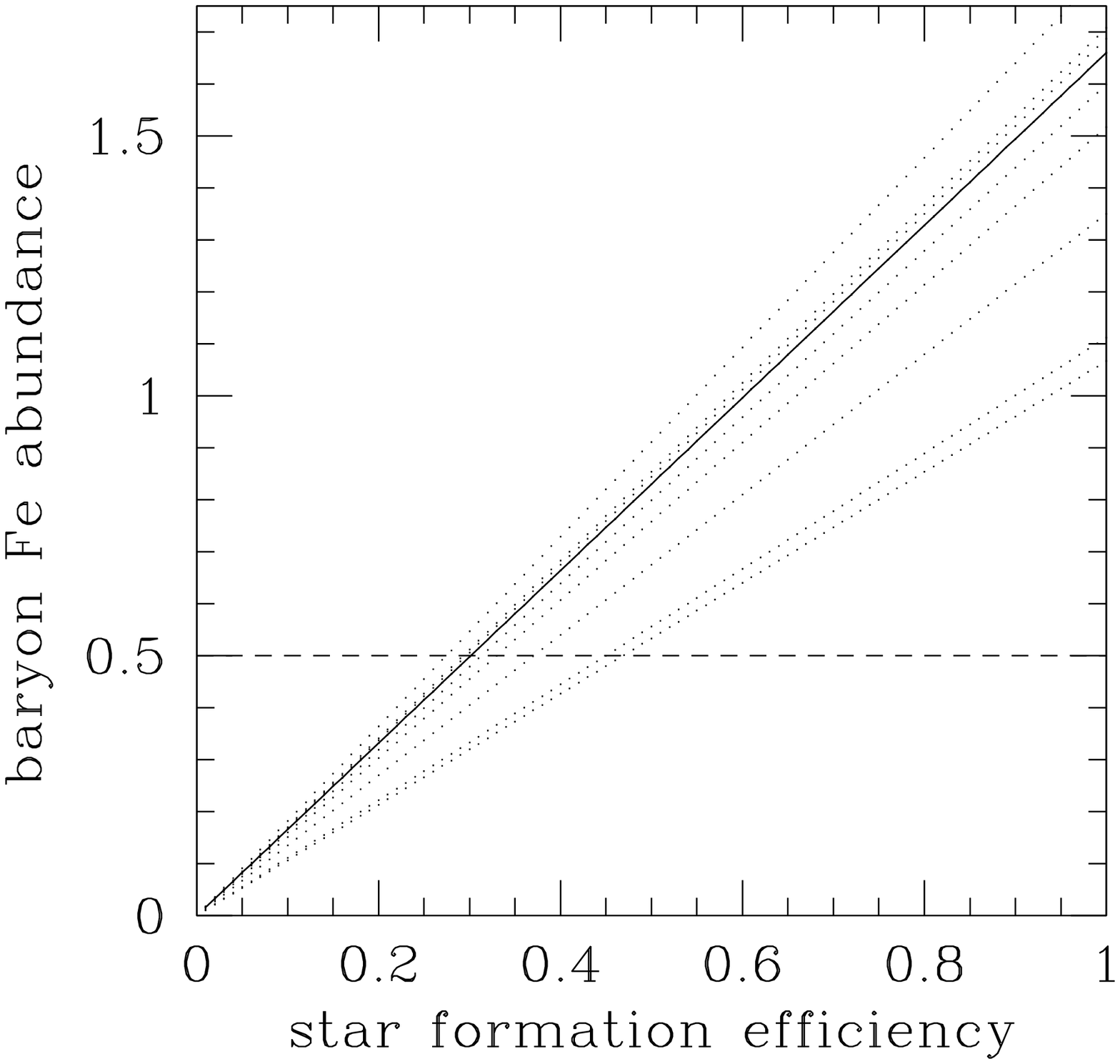}
\hfil
\includegraphics[scale=0.25,angle=0]{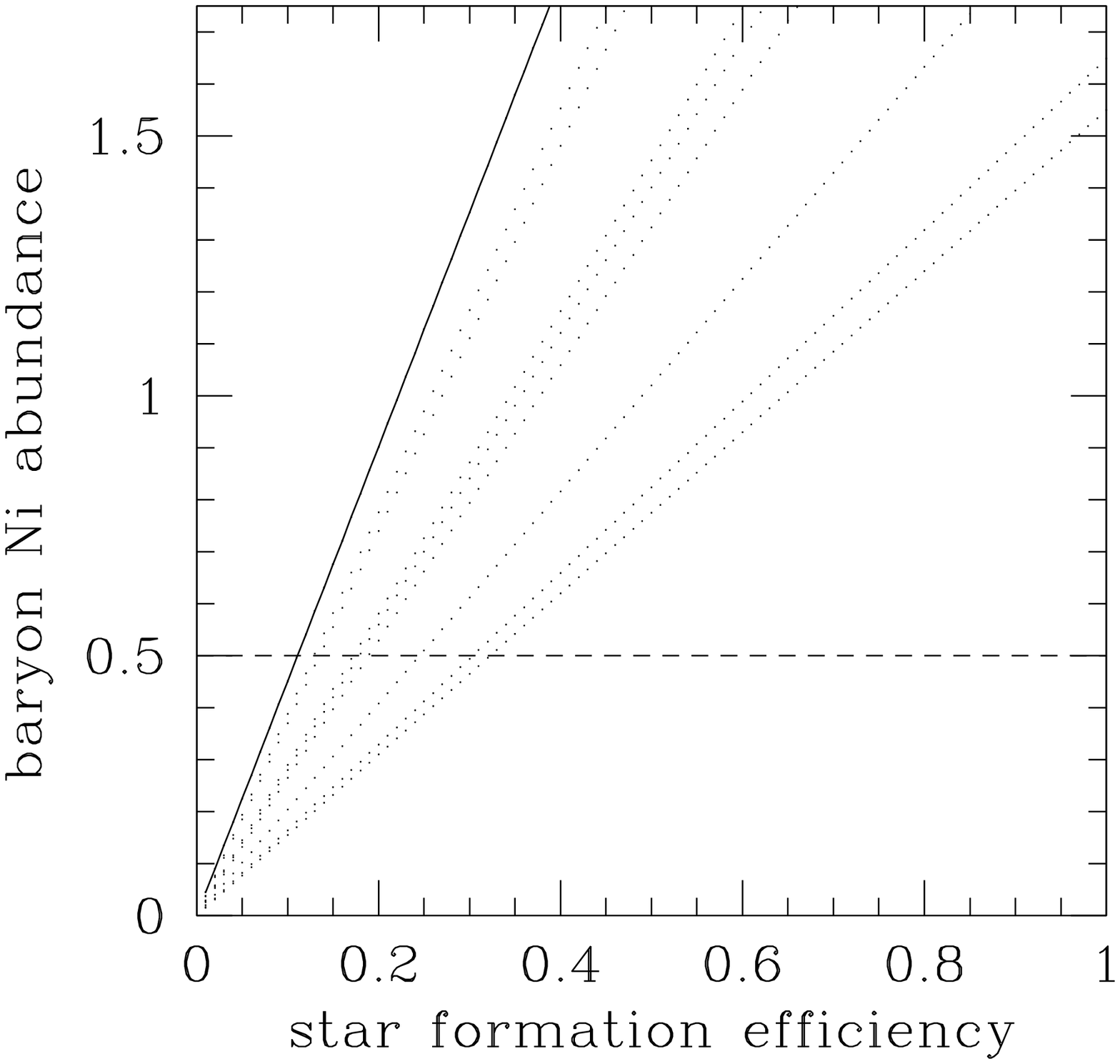}
\hfil
\includegraphics[scale=0.25,angle=0]{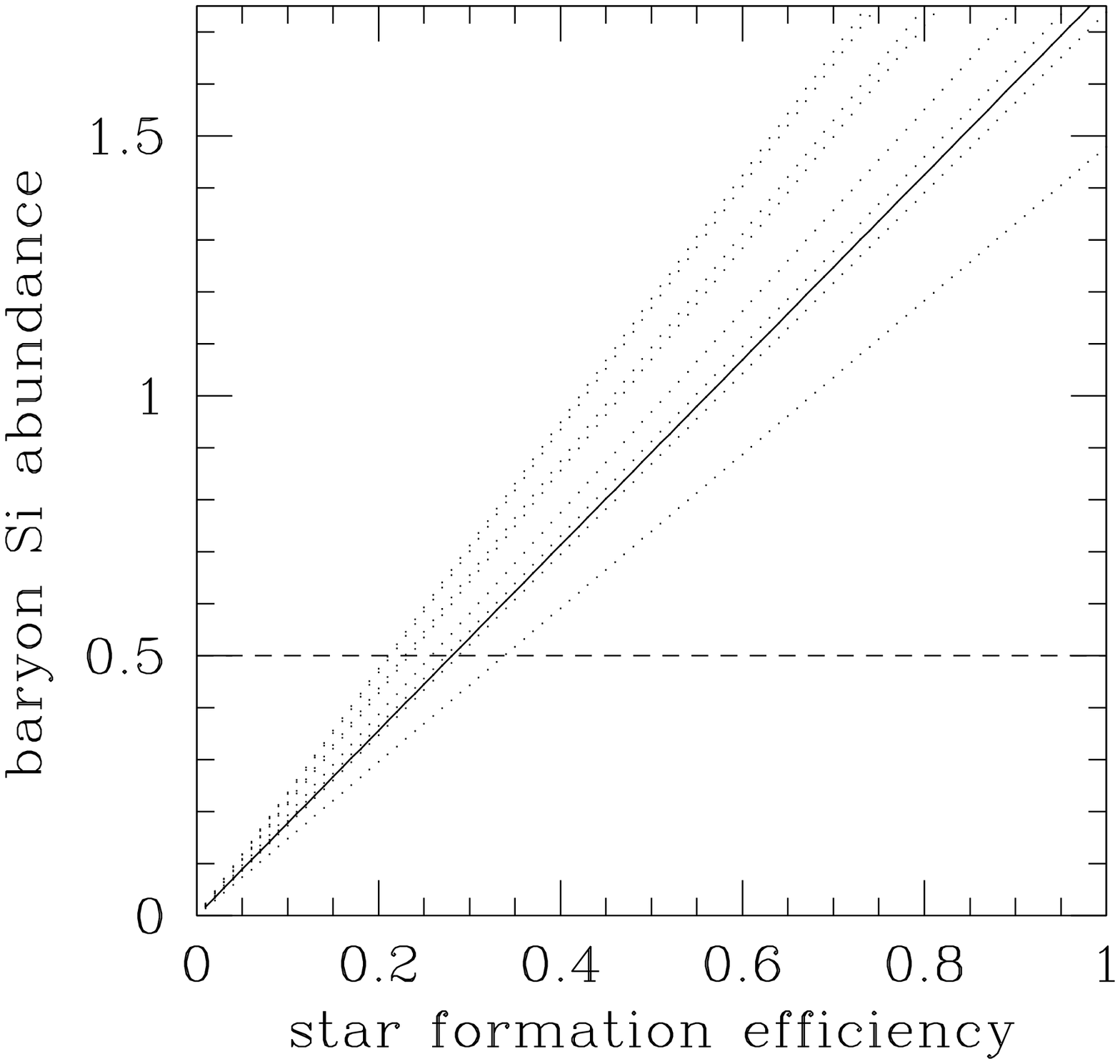}
\hfil
\caption{\footnotesize{{\it Left} panel (a): Average Fe abundance of
    all cluster baryons (stars and gas) as a function of star
    formation efficiency, $\varepsilon_{sf}$, assuming a single SNcc
    yield set but a wide range of SNIa yield sets (see text for
    details). The horizontal dotted line shows the very conservative
    value of $Z_{Fe,bar}=0.5$. {\it Middle} panel (b): Same as (a) for
    Ni. {\it Right} panel (c): Same as (a) for Si.}}
\end{figure}

\begin{figure}[h]
\includegraphics[scale=0.25,angle=0]{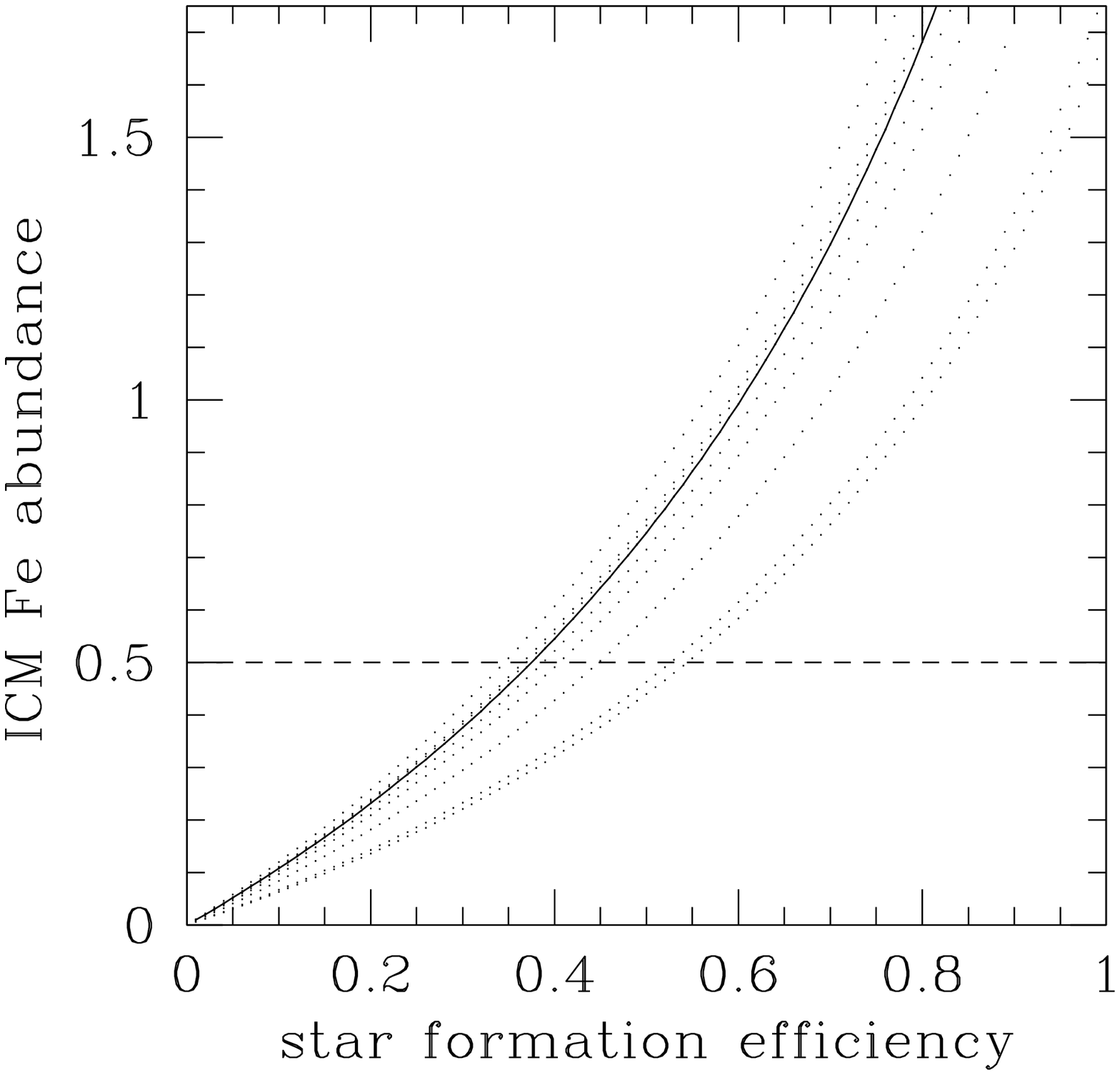}
\hfil
\includegraphics[scale=0.25,angle=0]{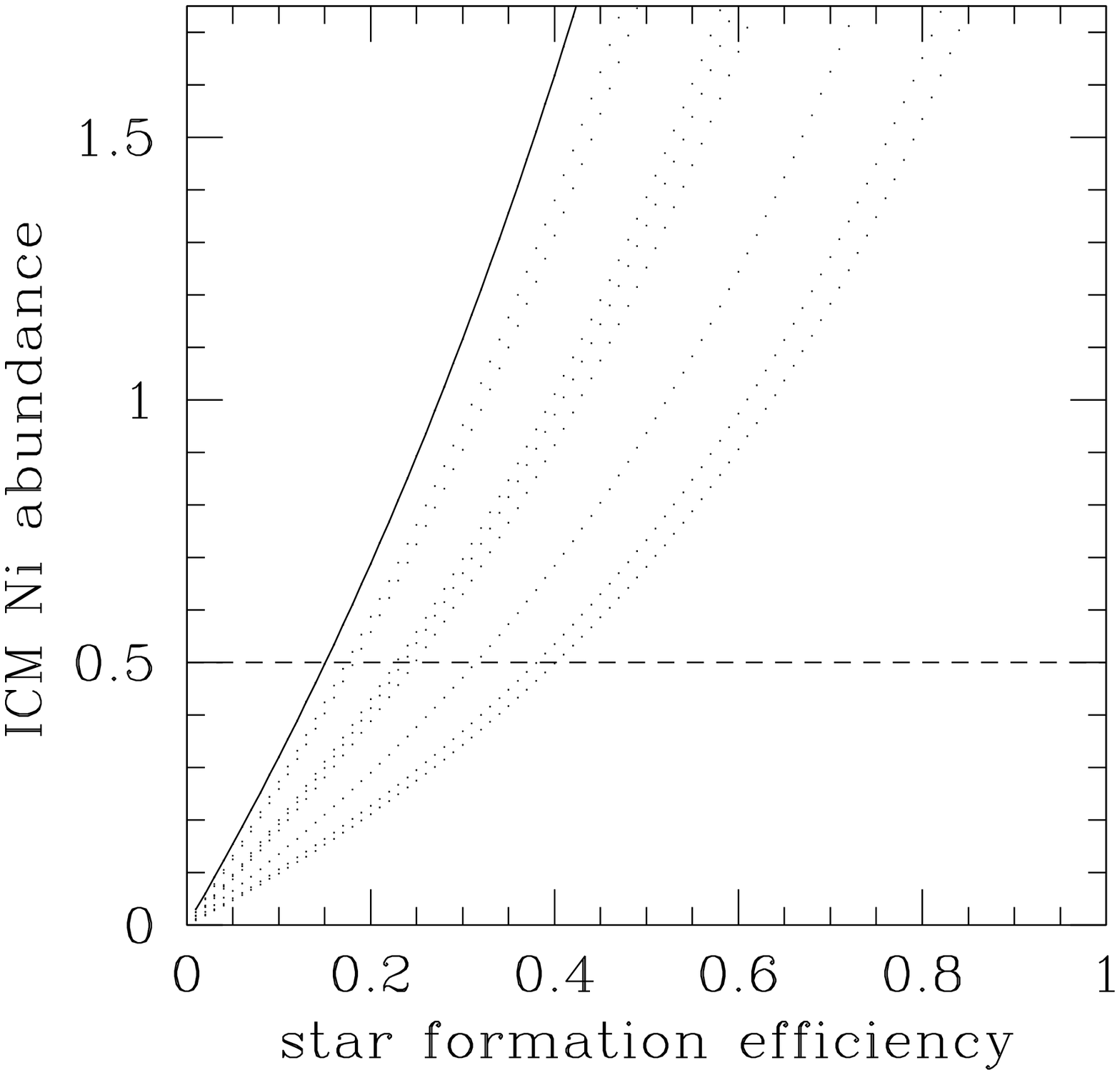}
\hfil
\includegraphics[scale=0.25,angle=0]{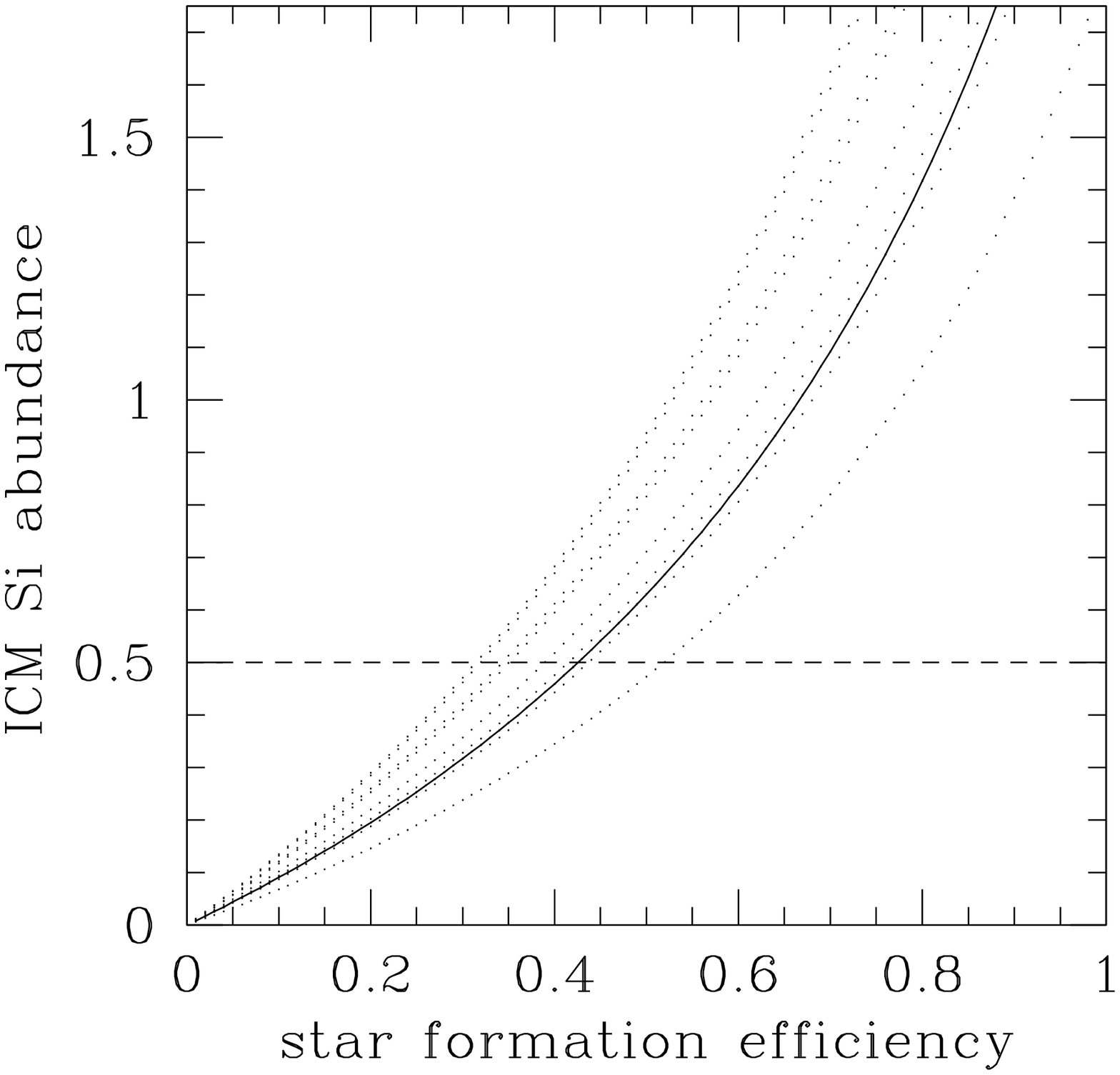}
\hfil
\caption{\footnotesize{Same as Figure 3 for ICM abundance, assuming
    the standard $\beta^{cc}=0.6$, $\alpha^{SN}=2$.}}
\end{figure}

\begin{figure}[h]
\includegraphics[scale=0.25,angle=0]{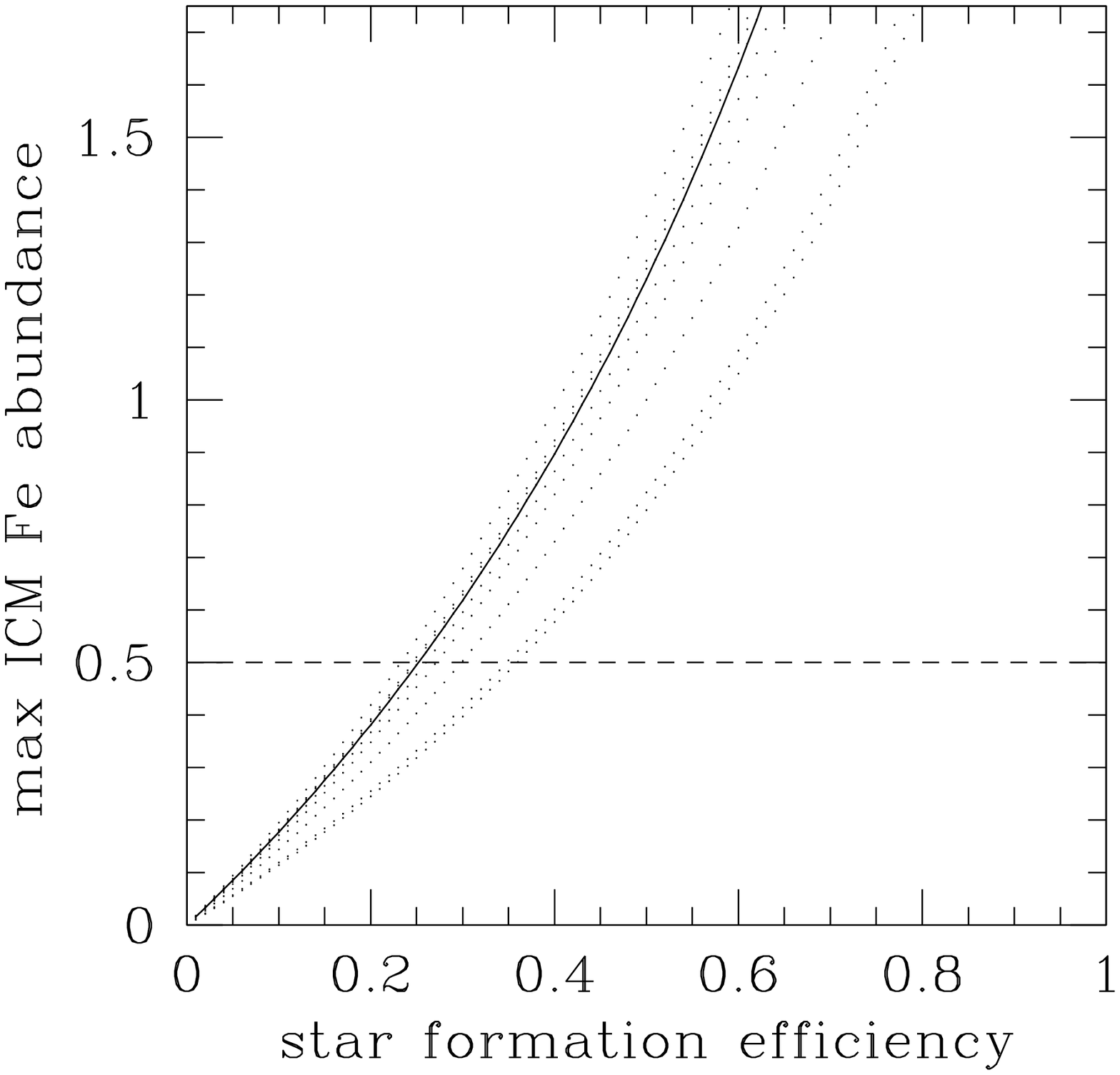}
\hfil
\includegraphics[scale=0.25,angle=0]{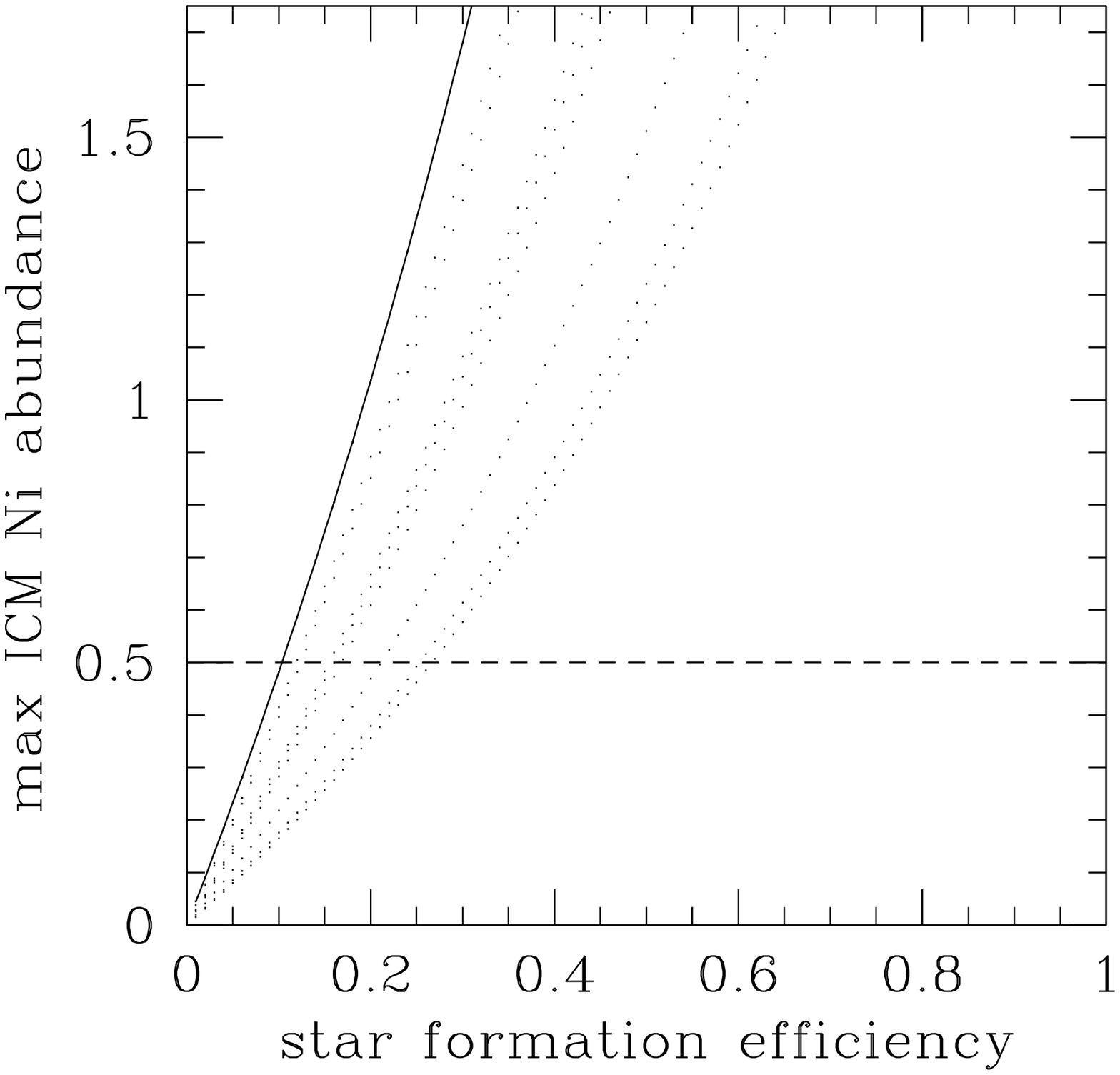}
\hfil
\includegraphics[scale=0.25,angle=0]{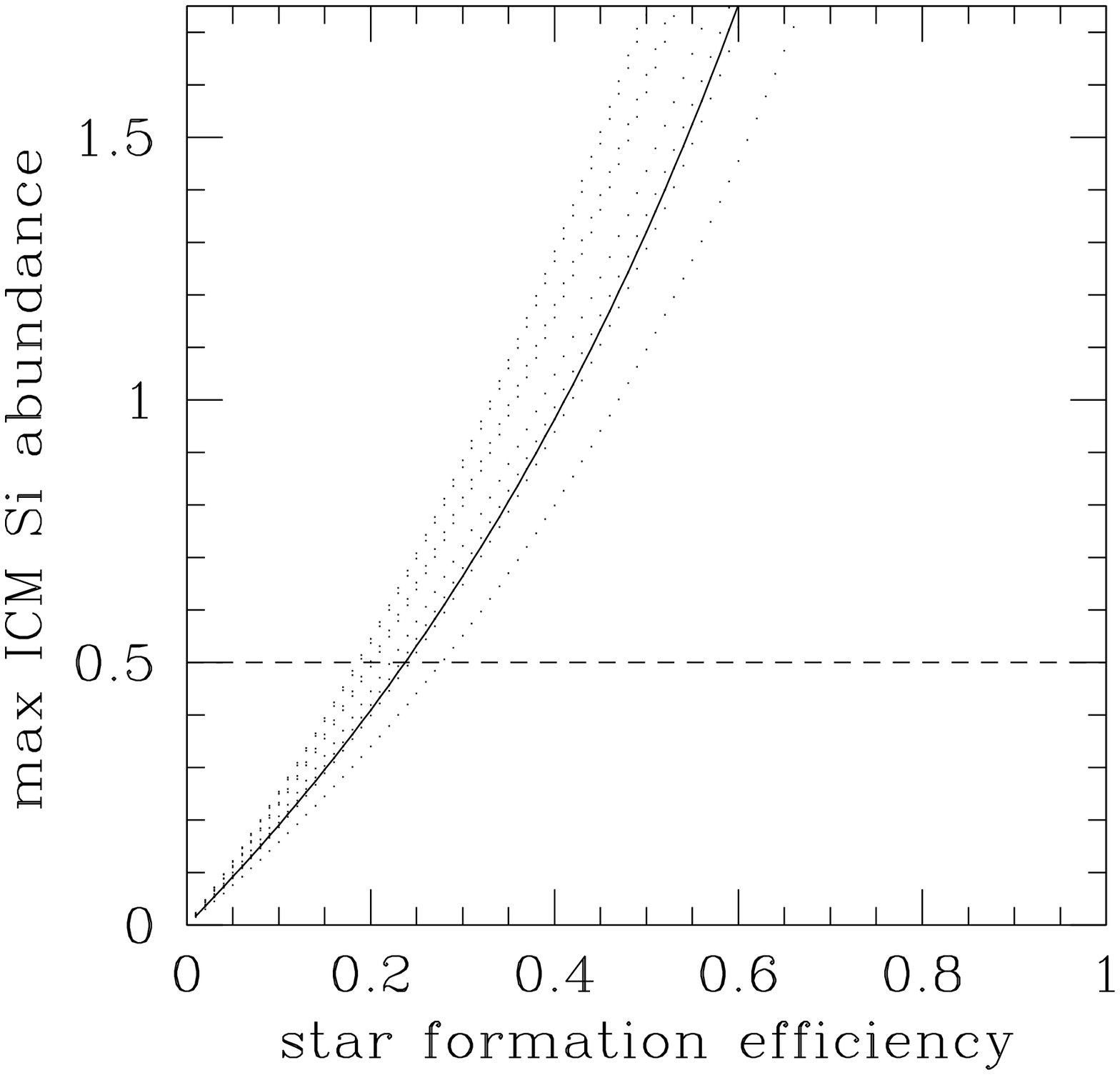}
\hfil
\caption{\footnotesize{Same as Figure 3 for the maximum ICM abundance,
    i.e. $\beta^{cc}=0$.}}
\end{figure}

From the results in this section, we confirm that ICM Fe abundances
cannot be produced if $\varepsilon_{sf}\sim 0.15$ ($f_{*}/f_{ICM}\sim
0.11$) and quantified the shortfall as a function of how efficiently
metals in general, and SNIa products in particular, are locked up in
stars. For $\varepsilon_{sf}\sim 0.3$ ($f_{*}/f_{ICM}\sim 0.24$), they
can -- but only for small lock-up fractions such that $\sim 85$\% of
SNIa, and $\sim 70$\% of SNcc, metal production is embedded in the ICM
and (by implication) $Z_{Fe,*}\sim 0.5$. The assumption of a standard
IMF is adopted throughout this section, an assumption we relax in the
following section.

\section{A More Comprehensive Examination (II): Effects of Changing the IMF}

The initial mass function (IMF) of stars in cluster galaxies and
intracluster space is intimately connected to estimates of ICM
enrichment, impacting the mass in stars calculated from the total
light, the mass return fraction (or, equivalently, the ratio of
current stellar mass to mass converted into stars), and the numbers of
SNcc and SNIa explosions expected per mass formed into stars. The
general characteristics of the IMF in various Milky Way
sub-populations are now well-determined and generally mutually
consistent \citep{bastian2010}. Although several functional forms are
commonly used for the ``canonical'' IMF \citep{kroupa2012}, these must
share the properties of a Salpeter-like slope at high mass with a
break to a flatter slope below $\sim 0.5-1~{\mathrm M}_{\sun}$. Many
of the best-studied environments are consistent with the hypothesis
that this form is ``universal'' in space and time, but variations in
extreme star forming environments that predominate in the early
universe (when most stars in the elliptical galaxies that dominate the
stellar content in clusters form) may explain a number of anomalies,
such as an apparent inconsistency between the observed evolution of
the global star formation rate and stellar mass densities (Narayanan
\& Dav\'e 2012a, and references therein).

Several recent observational investigations of elliptical galaxies
find direct evidence in the mass-to-light ratio for either an excess
of stellar remnants as realized in a ``top-heavy'' IMF; or, of low
mass stars as realized for a ``bottom-heavy'' IMF (e.g., Cappellari et
al. 2012; see Section 5 below). By exploiting the level and pattern of
ICM abundances we constrain the properties of the enriching stellar
population. We may then exclude particular elliptical galaxy IMFs
under the parsimonious assumption that these optically studied stars
originate from the same parent IMF as those that enrich the ICM -- or,
alternatively, call this assumption into question.

\subsection{Models and Parameters}

Simply put, the level of metal enrichment of the stellar and ICM
baryonic sub-components in clusters is a reflection of their
respective total masses, the total numbers of SNIa and SNcc that
enrich each constituent, and the nucleosynthetic yields of each of
these supernova explosion types. In previous sections, these are
expressed in terms of the mass return fraction, $r_{*}$, and formation
efficiency, $\varepsilon_{sf}$ of the stars, the total and relative
numbers of SNIa and SNcc per star formed ($\eta^{Ia}$, $R^{SN}$), and
phenomenological supernova lock-up and asymmetry parameters
($\beta^{cc}$ and $\alpha^{SN}$) characterizing the ultimate
destination (ICM or stars) of supernova products. In Appendix A these
are further deconstructed into more fundamental astrophysical
functions and parameters directly connected to stellar and galaxy
evolution, thus providing a self-consistent astrophysical framework
for understanding the effect of varying the IMF on cluster
enrichment. Ultimately these are reduced to the following: (1) The
functional form of the IMF (see below); (2) the present-day main
sequence turnoff mass ($0.9~{\mathrm M}_{\sun}$); (3) the
remnant-progenitor mass relationships for ($<8~{\mathrm M}_{\sun}$)
intermediate- (equation A4) and ($\ge8~{\mathrm M}_{\sun}$) high-mass
(equation A29) stars derived, for the former, from well-established
standard white dwarf masses and, for the latter, from a model for the
evolution of the stellar population and the delayed-explosion compact
remnant prescription in \cite{fryer2012}; (4) the ratio of mass
ejected from galaxies into the ICM during star formation to the mass
of stars formed, $\delta_{GW}$; (5) the galaxy formation efficiency,
$\varepsilon_{gal}$;\footnote{Defined as the fraction of baryons
  initially in galaxies, this is related to the star formation
  efficiency defined in equation (2) by the expression
  $\varepsilon_{sf}=\varepsilon_{gal}(1+\delta_{GW})^{-1}$.} and, (5)
various supernovae switches and parameters that we now describe. For
SNcc we assume progenitor masses from $m_{cc}=8~{\mathrm M}_{\sun}$ to
$m_{up}$, where $m_{up}$ may differ from the IMF upper mass limit
$m_{hi}$ (but is the same by default, and assumed so in calculating
the high mass return fraction). Since we adopt the IMF-averaged SNcc
yields of \cite{kobayashi2006} as a function of progenitor metallicity
$Z_{cc}$, $Z_{cc}$ must be specified as well. For SNIa we consider the
yield sets described in Section 3, assume progenitors in the
$3-8~{\mathrm M}_{\sun}$ range, and must specify the efficiency
$\varepsilon^{Ia}$ defined as the fraction of $3-8~{\mathrm M}_{\sun}$
that result in SNIa. In addition, the ``prompt'' fraction of SNIa that
explode during the star formation epoch and so may be incorporated
into stars or ICM, $f^{Ia}_p$, (while a fraction $1-f^{Ia}_p$ strictly
enrich the ICM) must be specified.

Our default IMF is the \cite{kroupa2012} segmented power-law with
slopes and mass scales that can explain local star formation (Appendix
A); other defaults are $Z_{cc}=1$, W7 yields, $\varepsilon_{gal}=0.25$
and $\delta_{GW}=0.5$ ($\varepsilon_{sf}=0.17$, $f_{*}/f_{ICM}\sim
0.11$), $\varepsilon^{Ia}=0.076$ (Section A.2), and
$f^{Ia}_p=0.5$. Under these conditions, $r_{*}=0.41$ ($r^{im}_*=0.25$,
$r^{hi}_*=0.16$ for intermediate- and high- mass stars, as delineated
above) while the fraction in stellar remnants is 0.17 (0.11/0.06 from
intermediate/high-mass stars), $\eta^{cc}=0.011$, $\eta^{Ia}=0.0022$,
$\alpha^{SN}=2$ and $\beta^{cc}=0.39$ -- similar to the default
parameters in Section 3. The resulting abundances are shown in Table 1
where, once again as expected, we find reasonable stellar abundances
but ICM abundances too low by a factor of $\sim 2$. For comparison we
also display the abundances for a Salpeter IMF, which fails to provide
sufficient metals for {\it all} components (including stars) for this
default set of parameters.\footnote{It should be noted that most
  observational estimates of $f_{*}/f_{ICM}$ are IMF-dependent, and
  would be larger for a Salpeter IMF given a fixed amount of optical
  light.}

\begin{deluxetable}{cccccc}
\tabletypesize{\scriptsize} \tablewidth{0pt} \tablecaption{ICM
  Abundances from Default Parameters} \tablehead{\colhead{} &
  \colhead{0} & \colhead{Mg} & \colhead{Si} & \colhead{Fe} &
  \colhead{Ni}}
\startdata Baryons & 0.44 (0.27) & 0.30 (0.18)& 0.38 (0.23) & 0.32
(0.20) & 0.84 (0.53)\\ Stars & 1.74 (1.05) & 1.19 (0.72) & 1.35 (0.82)
& 0.87 (0.54) & 1.90 (1.21)\\ ICM & 0.30 (0.15) & 0.20 (0.10) & 0.27
(0.15) & 0.26 (0.15) & 0.72 (0.43)\\ \enddata
\tablecomments{All abundances relative to \citep{asplund2009} solar
  standard. The values in parentheses are for a Salpeter IMF with the
  default values of $\varepsilon_{gal}$, $\delta_{GW}$,
  $\varepsilon^{Ia}$, and $f^{Ia}_p$, and the same range of masses
  ($0.07-150~{\mathrm M}_{\sun}$).}
\end{deluxetable}

\begin{deluxetable}{ccccc}
\tabletypesize{\scriptsize} \tablewidth{0pt} \tablecaption{Model IMF
  Parameters} \tablehead{\colhead{} & \colhead{$m_{lo}$} &
  \colhead{$m_{br}$} & \colhead{$\alpha_{lo}$} &
  \colhead{$\alpha_{hi}$}} \startdata {\it sl-0} & 0.07 & \nodata &
$1.8\rightarrow 2.3$ & $=\alpha_{lo}$ \\ {\it sl-1} & 0.07 & 0.5 &
$0.3\rightarrow 2.4$ & 2.3 \\ {\it sl-2} & 0.07 & 1.0 &
$0.3\rightarrow 2.3$ & 2.3 \\ {\it sl-3} & 0.07 & 0.5 & 1.3 &
$1.8\rightarrow 2.5$ \\ {\it sl-4} & 0.07 & 1.0 & 1.3 &
$1.8\rightarrow 2.5$ \\ {\it sl-5} & 0.07 & 8.0 & 1.3 &
$1.5\rightarrow 3.3$ \\ {\it m-1} & $0.01\rightarrow 1.0$ & 1.0 & 1.3
& 2.3 \\ {\it m-2} & $0.01\rightarrow 0.5$ & 0.5 & 1.3 & 2.3 \\ {\it
  m-3} & $0.07$ & $0.5\rightarrow 5.45$ & 1.3 & 2.3 \\ \enddata
\tablecomments{$m_{hi}=150~{\mathrm M}_{\sun}$ for all models
  displayed here. $m_{br}$ -- equivalent to $m2=0.5~{\mathrm
    M}_{\sun}$ for models {\it sl-1}, {\it sl-3}, and {\it m-2}; and
  $m_3$ (default: $1~{\mathrm M}_{\sun}$) for models {\it sl-2}, {\it
    sl-4}, {\it sl-5}, {\it m-1}, and {\it m-3} -- is defined as the
  mass where the IMF slope transitions to its high-mass value. Models
  {\it sl-4} and {\it sl-5} include an additional break from
  $\alpha_1=1.3$ to $\alpha_2=2.3$ at $m_2$.  $\alpha_{1}=\alpha_{lo}$
  and $\alpha_{3}=\alpha_{hi}$, while $\alpha_{2}=\alpha_{lo}$ for
  models {\it sl-2}, {\it m-1}, and {\it m-3}; $\alpha_{2}=\alpha_{3}$
  for models {\it sl-1}, {\it sl-3}, {\it m-2}; and, is set at the
  default $\alpha_{2}=2.3$ for models {\it sl-1}, {\it sl-4}, {\it
    sl-5}, and {\it m-2}. Parameter ranges correspond to those with
  physical solutions; i.e., $f^{Ia}_p$ and $\delta_{GW}>0$ and
  $\varepsilon_{gal}<1$ -- for the {\it sl-4} model with higher SNIa
  efficiency (star-to-gas ratio) the range shifts to
  $1.5\rightarrow2.5$ ($2\rightarrow2.5$); see below and Tables 3 and
  4.}
\end{deluxetable}

Our approach to examining the effects of varying the IMF on ICM
enrichment, that attempts to make comparisons at fixed values of the
observables to the extent possible, is as follows. As detailed in
Appendix A, we consider departures from the ``canonical'' IMF (Table
2) described according to sequences with either a single slope below
the break mass $m_3$, or an additional distinct slope below
0.5~${\mathrm M}_{\sun}$. Either the lower mass limit ($m_{lo}$), or
the slopes ($\alpha_1$ or $\alpha_3$) at low or high mass ends may be
varied. Additionally, $m_3$ may vary in sequences of the first
type. To isolate the effect of the IMF on ICM abundances, we impose
invariance on the stellar Fe and O abundances (Table 1). This
determines the necessary adjustments in the parameters $\delta_{GW}$
and $f^{Ia}_p$ (for some $\varepsilon^{Ia}$). For fixed yield sets,
invariance in $Z_{Fe,*}$ and $Z_{O,*}$ assures invariance in all
stellar abundances. Finally, we consider these variations at fixed
{\it present-day} baryon inventory (11\% stars, 89\% ICM), which is
equivalent to adjusting $\varepsilon_{gal}$ so as to maintain constant
$\varepsilon_{sf}(1-r_*)$ (equations 2, A26) -- thus enabling us to
investigate what adjustments in IMF (if any) may explain observed ICM
abundances assuming this nominal star-to-gas ratio. The imposition of
these constraints rule out those IMFs that imply unphysical values of
$\delta_{GW}$($<0$), $f^{Ia}_p$($>1$), or $\varepsilon_{gal}$($>1$) --
i.e., some IMFs are incompatible with observed stellar abundances and
a $\sim$9:1 ratio of ICM to stars (thus the limited range of the
variable IMF parameters in Table 2).

\subsection{Impact of Varying the IMF}

We remind the reader that our standard IMF has slope $\alpha=1.3$
below 0.5~${\mathrm M}_{\sun}$, and 2.3 above. Figures 6a-d show the
impact on ICM enrichment of varying one (and only one) of the slopes
and (in some cases) adjusting the break mass, by plotting the Fe
abundance and Mg/Fe, Si/Fe, and Ni/Fe ratios versus the deviation in
the non-fixed slope from these standard values.\footnote{Mg is almost
  exclusively synthesized in SNcc, Si primarily (but not exclusively)
  in SNcc, Fe in both SNcc and SNIa, and Ni primarily in SNIa.} This
covers many of the IMFs considered in the literature as possibly
resolving various conflicts between expectations and observations of
stellar populations in elliptical and/or starburst galaxies. For
models {\it sl-1} ({\it sl-2}) $\alpha$ below 0.5 ${\mathrm M}_{\sun}$
(1~${\mathrm M}_{\sun}$) is varied with $\alpha$ above these single
break masses fixed at 2.3 -- {\it i.e.} positive (negative)
$\Delta\alpha$ corresponds to bottom-heavy (-light) IMFs. For models
{\it sl-3}, {\it sl-4}, and {\it sl-5}, $\alpha$ is varied above break
masses of 0.5 ${\mathrm M}_{\sun}$, 1~${\mathrm M}_{\sun}$, and
8~${\mathrm M}_{\sun}$, respectively, with $\alpha=2.3$ between 0.5
${\mathrm M}_{\sun}$ and the break mass in the latter two. For these
models, positive (negative) $\Delta\alpha$ corresponds to top-light
(-heavy) IMFs. In addition, we plot the results for a single-slope IMF
({\it sl-0}). We can see that $Z_{Fe,ICM}\sim 0.5$ is predicted for
either a (1) bottom-light IMF with $\alpha(\le 1{\mathrm
  M}_{\sun})\sim 1$, (2) top-heavy IMF with $\alpha(>0.5{\mathrm
  M}_{\sun})\sim 2$ or $\alpha(>1{\mathrm M}_{\sun})\sim 1.8$, (3)
single-slope IMF ({\it i.e.} both bottom- and top-heavy) with
$\alpha\sim 1.8$. As expected, $Z_{Fe,ICM}\sim 0.1$ for the pure
Salpeter IMF (single slope and bottom-light cases with slope 2.35).


\begin{figure}[h]
\includegraphics[scale=0.8,angle=0]{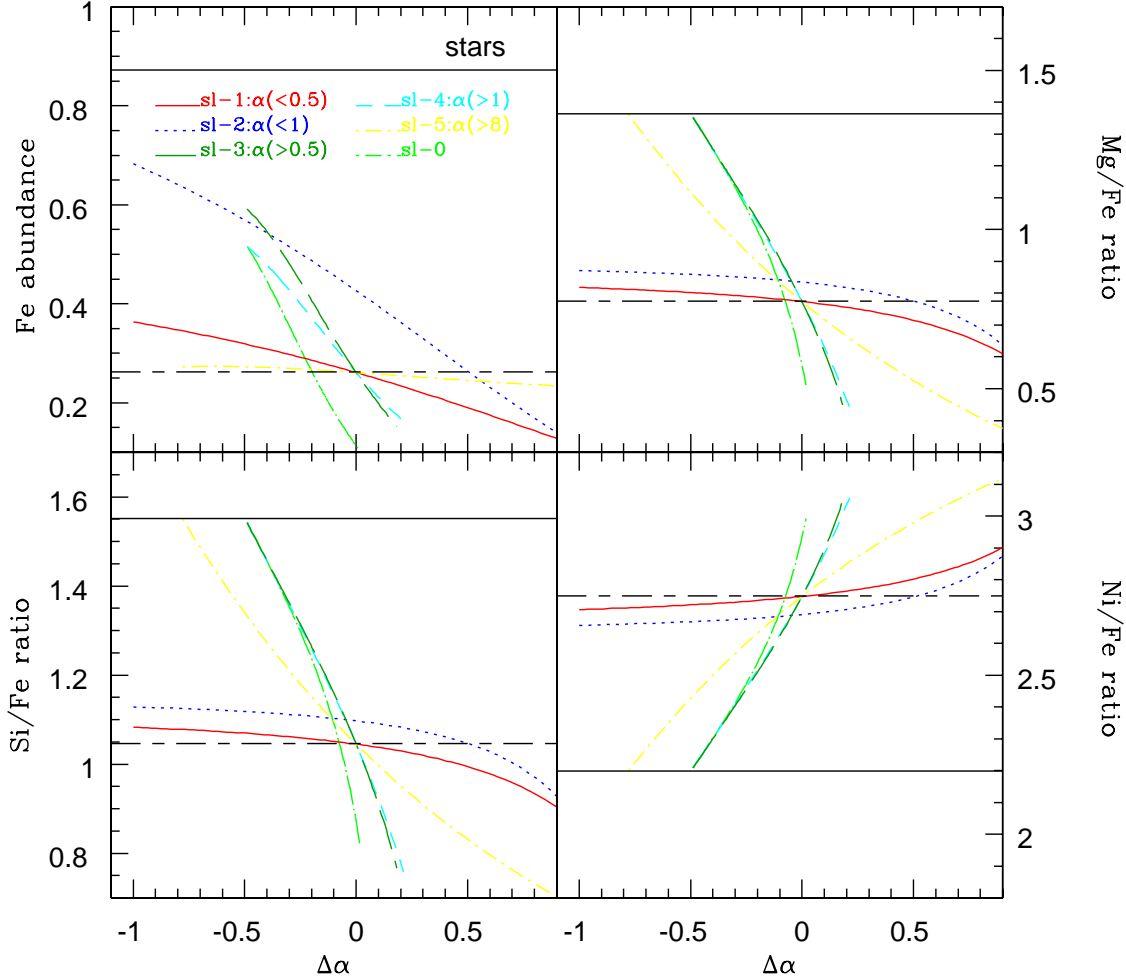}
\caption{\footnotesize{{\it Clockwise from upper left} (a)-(d): ICM Fe
    abundance and Mg/Fe, Ni/Fe, and Si/Fe abundance ratios for models
    with departures, with respect to the standard IMF, in either high-
    or low-mass slope: $\Delta\alpha=\alpha_{hi}-2.3$ for
    top-heavy/light ({\it sl-1}, {\it sl-2}) and $\alpha_{lo}-1.3$ for
    bottom-light/heavy ({\it sl-3}, {\it sl-4} , {\it sl-4}), models
    -- see text and Table 2 for details. Curves for models {\it sl-3}
    and {\it sl-4} closely trace each other in the ratio plots. The
    solid and broken horizontal lines, respectively, show the stellar,
    and standard model ICM, values (Table 1).}}
\end{figure}

Varying the lower mass cutoff provides an alternative means of
producing either bottom-heavy ($m_{lo}<0.07~{\mathrm M}_{\sun}$) or
bottom-light ($m_{lo}>0.07~{\mathrm M}_{\sun}$) IMFs, while increasing
the break mass that delineates $\alpha=1.3$ from $\alpha=2.3$
\citep{narayanan2012a} is also bottom light in the sense that the
fraction of low-mass stars relative to those at intermediate and high
mass is suppressed. As shown in Figure 7a, these alternatives can
explain $Z_{Fe,ICM}=0.5$ for IMFs with the standard high mass slope
$\alpha=2.3$ if the IMF lower mass cutoff is shifted from the standard
$m_{lo}=0.07~{\mathrm M}_{\sun}$ to $m_{lo}\sim 0.2~{\mathrm
  M}_{\sun}$ for an IMF with a single break at 1~${\mathrm M}_{\sun}$,
what is (essentially) a Salpeter IMF with $m_{lo}\sim 0.5~{\mathrm
  M}_{\sun}$, or for an IMF with the break between $\alpha=1.3$ and
$\alpha=2.3$ shifted from 0.5 to $\sim 1~{\mathrm M}_{\sun}$. The last
would seem to represent a particularly modest departure from the
standard IMF.


\begin{figure}[h]
\includegraphics[scale=0.8,angle=0]{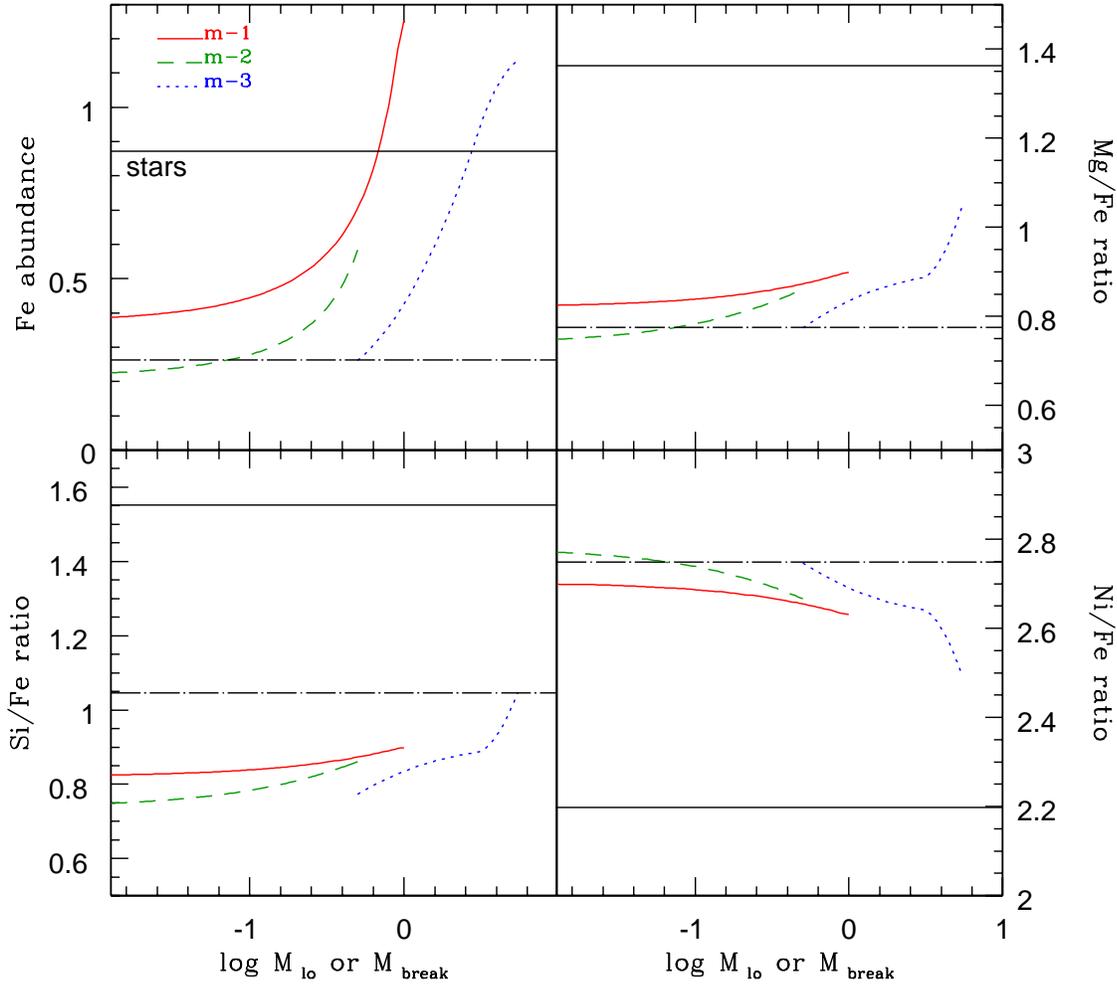} 
\caption{\footnotesize{Same as figure 6 for departures, with respect to
    the standard IMF, in lower mass limit or break mass -- see text
    and Table 2 for details.}}
\end{figure}

Bottom-light and top-heavy scenarios may be directly distinguished in
ICM spectra via abundance ratios, as demonstrated in Figures 6b-d,
7b-d, and 8a-c; and Table 3. These essentially define two branches in
the $Z_{Fe,ICM}-(\alpha/Fe)_{ICM}$ plane (Figure 8), with ratios in
the top-heavy branch connecting to the stellar ratio as the IMF
flattens (and $f^{Ia}_p\rightarrow 1$), and abundances for the
bottom-light IMFs (assured to have $f^{Ia}_p=0.5$) universally rising
in lockstep with the elimination of low-mass stars. Table 3, confined
to those models that predict $Z_{Fe,ICM}=0.5$, illustrates how the
$\alpha/Fe_{Fe,ICM}$ ratios might be exploited to distinguish among
bottom-light and top-heavy IMF explanations for ICM enrichment -- for
the former, ratios deviate more strongly from those in stars (smaller
$\alpha/Fe_{Fe,ICM}$, larger Ni/Fe) and provide a better match to the
data \citep{simionescu2009}.

\begin{figure}[h]
\includegraphics[scale=0.25,angle=0]{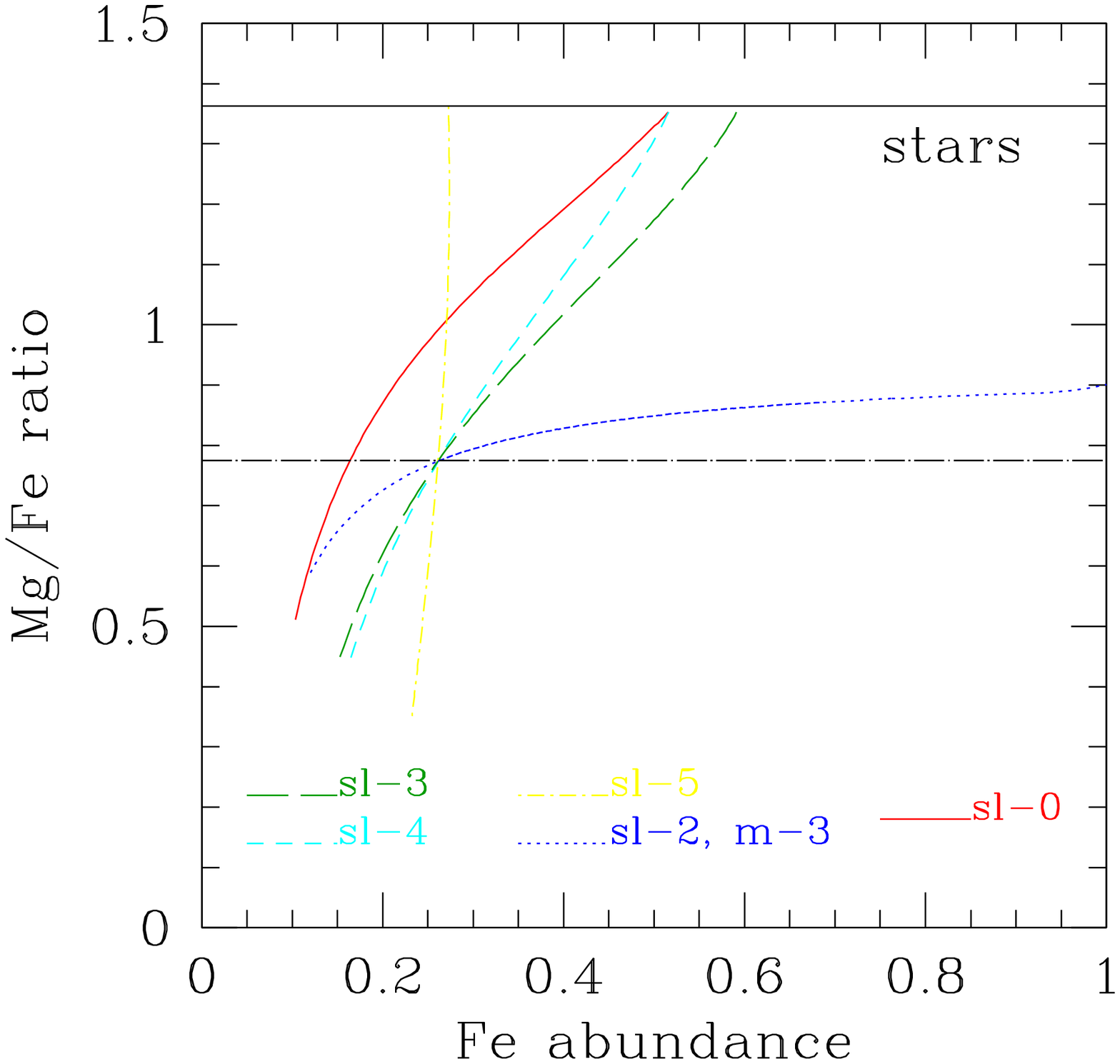}
\hfil
\includegraphics[scale=0.25,angle=0]{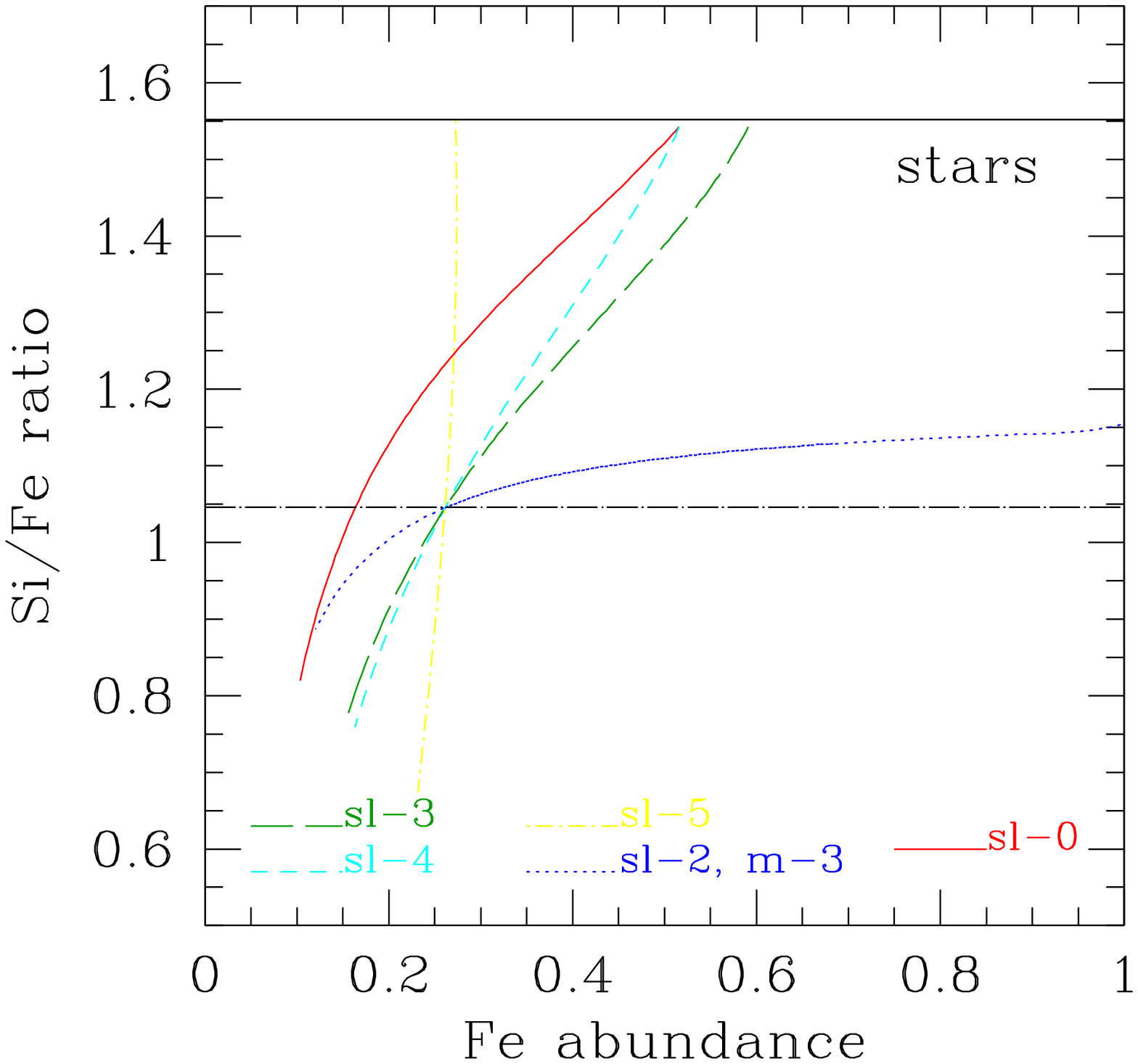}
\hfil
\includegraphics[scale=0.25,angle=0]{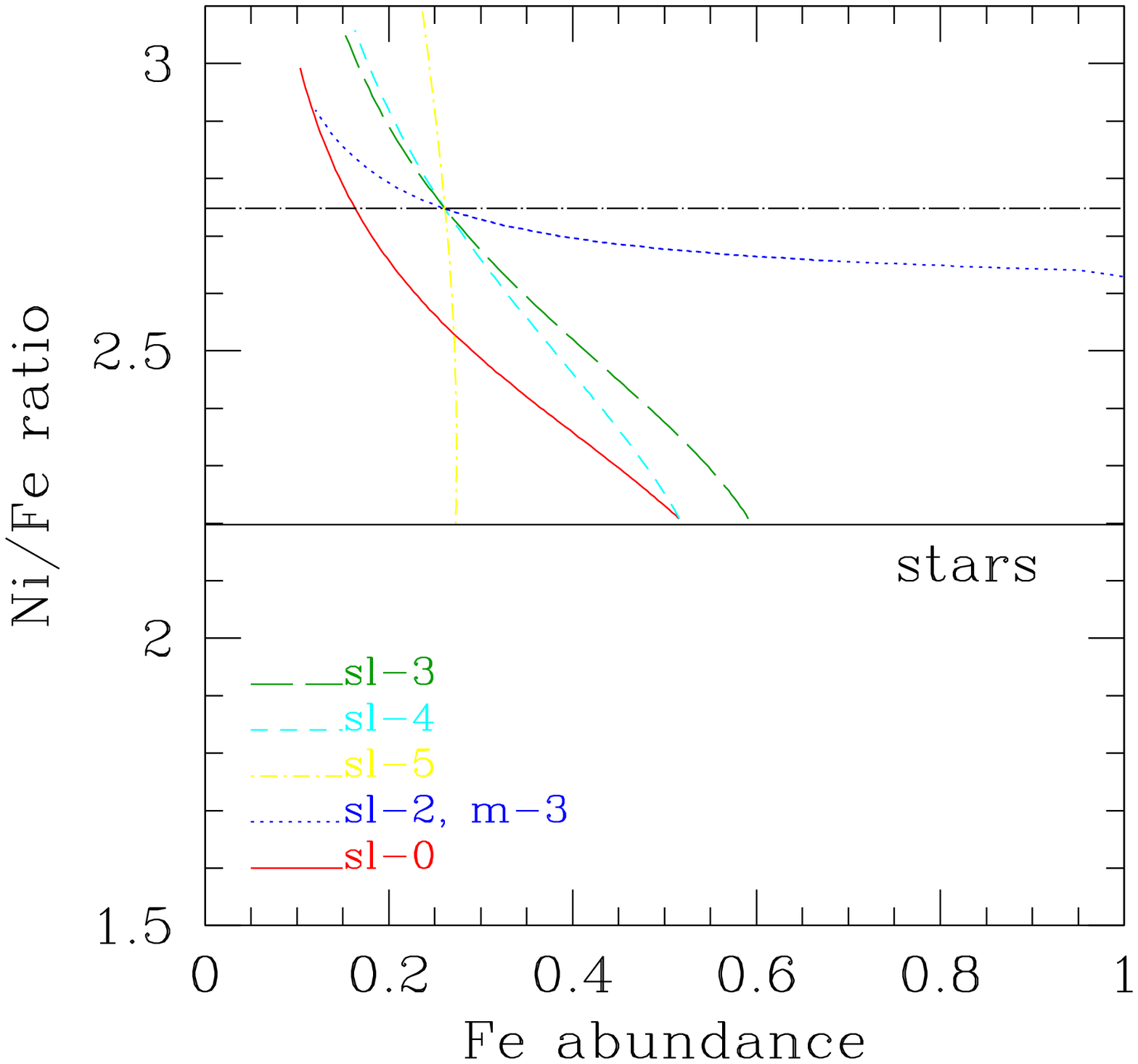}
\hfil
\caption{\footnotesize{ICM abundance ratios, with respect to Fe, for
    Mg -- {\it left} panel (a), Si -- {\it middle} panel (b), and Ni
    -- {\it right} panel (c) -- {\it vs.} Fe abundance for top-heavy
    models {\it sl-0}, {\it sl-3}, {\it sl-4}, and {\it sl-5}; and,
    for bottom-light models {\it sl-2} and {\it m-3} (all bottom-light
    models considered essentially follow the same curve).}}
\end{figure}

\begin{deluxetable}{ccccccc}
\tabletypesize{\scriptsize} \tablewidth{0pt} 
\tablecaption{Abundance Ratios in Models with $Z_{Fe,ICM}=0.5$} 
\tablehead{\colhead{} & \colhead{} & 
\colhead{$O/Fe$} & \colhead{$Mg/Fe$} & \colhead{$Si/Fe$} &
\colhead{$Ni/Fe$} } \startdata
{\it stars} & \nodata & 2.0 & 1.4 & 1.6 & 2.2 \\
{\it salpeter} & \nodata & 1.0 & 0.68 & 0.97 & 2.8 \\
{\it canonical} & \nodata & 1.1 & 0.77 & 1.0 & 2.7 \\
{\it sl-0} & $\alpha=1.83$ & 1.9 & 1.3 & 1.5 & 2.2 \\
{\it sl-2} & $\alpha_{lo}=1.06$ & 1.2 & 0.85 & 1.1 & 2.7 \\
{\it sl-3} & $\alpha_{hi}=1.97$ & 1.7 & 1.2 & 1.4 & 2.4 \\
($\varepsilon^{Ia}=0.13$) & $\alpha_{hi}=2.22$ & 0.85 & 0.57 & 0.87 & 2.9 \\
(star/gas=0.25) & $\alpha_{hi}=2.36$ & 1.0 & 0.69 & 0.97 & 2.8 \\
{\it sl-4} & $\alpha_{hi}=1.85$ & 1.9 & 1.3 & 1.5 & 2.3 \\
{\it m-1} & $m_{lo}=0.20$ & 1.2 & 0.85 & 1.1 & 2.7 \\
{\it m-2} & $m_{lo}=0.42$ & 1.2 & 0.85 & 1.1 & 2.7 \\
{\it m-3} & $m_{br}=1.25$ & 1.2 & 0.85 & 1.1 & 2.7 \\ 
\enddata 
\tablecomments{Canonical and Salpeter model ratios, and stellar
  ratios, included for comparison purposes; $\varepsilon^{Ia}=0.13$
  and star/gas=0.25 variations of model {\it sl-3} also included.}
\end{deluxetable}

\subsection{Implications of Models with Nonstandard IMFs}

Table 4 displays the essential characteristics of models that produce
$Z_{Fe,ICM}=0.5$, and are constrained to match the standard values of
stellar metallicity and $f_{*}/f_{ICM}$.\footnote{Effects of relaxing
  the latter are discussed shortly.} Several general properties, as
well as others that distinguish top-heavy from bottom-light solutions
emerge. Relative to the model with canonical IMF, all have a relative
deficiency of unevolved low-mass stars, and hence $\sim 30$\% higher
mass return ($r_{*}$) and remnant ($f_{rem}$) fractions -- implying
$\sim 30$\% upward adjustments in the integrated mass of stars formed
based on the present-day mass, and $\sim 60$\% upward adjustments
based on the luminous stellar mass. The ``extra'' metals are explained
by a larger fraction of stars in the supernova-progenitor mass range,
and a larger ratio of mass in stars formed to present-day stellar
mass.

Successful top-heavy models are characterized by a large ``prompt''
fraction of SNIa and prodigious galactic winds, and have low lock-up
fraction and modest asymmetry between stellar and ICM abundance
patterns. That is, the extra ICM metals are mostly associated with the
rapid early star formation epoch where both formation of SNcc and SNIa
progenitors, and delivery of the metals from star forming sites to
extragalactic hot gas, are efficiently realized.

Successful bottom-light models are characterized by the standard
prompt SNIa fraction and more modest (though still substantial)
galactic winds, with a larger fraction of ICM enrichment occurring
during the passive, post-star-formation phase. In these models, the
implied fraction of mass originally in galaxies is 0.42 and the
fraction of the ICM that is primordial, $f_{prim}=0.65$ -- compared to
0.25 and 0.83 in the canonical model. For the top-heavy models these
take on more extreme values -- each on the order 0.5 -- although,
due to the effects of galactic winds, the star formation efficiencies
are not appreciably different.  

The different abundance patterns predicted in bottom-light and
top-heavy models is reflected in values of $R^{SN}_{ICM}$ that are
$\sim$twice as high for the former, more consistent with -- though
still lower than -- the value recently inferred by \citep{bsl2012} for
the central region in Abell 3112.

\begin{deluxetable}{ccccccccccccccc}
\tabletypesize{\scriptsize} \tablewidth{0pt} 
\tablecaption{Characteristics of Models with $Z_{Fe,ICM}=0.5$} 
\tablehead{\colhead{} & \colhead{} & \colhead{$r_{*}$} &
  \colhead{$f_{rem}$} & \colhead{$f^{Ia}_p$} & \colhead{$\delta_{GW}$}
  &\colhead{$\varepsilon_{gal}$} & \colhead{$\varepsilon_{sf}$} &
  \colhead{$\eta^{Ia}$} &\colhead{$\eta^{cc}$} &
  \colhead{$\beta^{cc}$} & \colhead{$\alpha^{SN}$} &
  \colhead{$R^{SN}_{ICM}$} & \colhead{$f_{prim}$}}\startdata
{\it salpeter} & \nodata & 0.27 & 0.11 & 0.5 & 0.5 & 0.25 &
0.17 & 0.0014 & 0.0065 & 0.50 & 2.0 & 0.32 & 0.86\\
{\it canonical} & \nodata & 0.41 & 0.17 & 0.5 & 0.5 & 0.25 &
0.17 & 0.0022 & 0.011 & 0.39 & 2.0 & 0.27 & 0.83\\
{\it sl-0} & $\alpha=1.83$ & 0.56 & 0.24 & 0.96 & 1.7 & 0.60 &
0.22 & 0.0020 & 0.019 & 0.16 & 1.0 & 0.11 & 0.44\\\
{\it sl-2} & $\alpha_{lo}=1.06$ & 0.54 & 0.22 & 0.5 & 0.97 & 0.42 &
0.21 & 0.0028 & 0.014 & 0.24 & 2.0 & 0.23 & 0.65\\
{\it sl-3} & $\alpha_{hi}=1.97$ & 0.54 & 0.23 & 0.78 & 1.5 & 0.54 &
0.22 & 0.0023 & 0.018 & 0.18 & 1.3 & 0.13 & 0.50\\
($\varepsilon^{Ia}=0.13$) & $\alpha_{hi}=2.22$ & 0.45 & 0.18 & 0.33 &
0.75 & 0.31 & 0.17 & 0.0039 & 0.013 & 0.32 & 3.1 & 0.40 & 0.76\\
(star/gas=0.25) & $\alpha_{hi}=2.36$ & 0.38 & 0.16 & 0.46 &
0.33 & 0.43 & 0.32 & 0.0021 & 0.0096 & 0.46 & 2.2 & 0.32 & 0.72\\
{\it sl-4} & $\alpha_{hi}=1.85$ & 0.55 & 0.24 & 0.93 & 1.7 & 0.59 &
0.22 & 0.0021 & 0.019 & 0.17 & 1.1 & 0.11 & 0.45\\
{\it m-1} & $m_{lo}=0.20$ & 0.54 & 0.22 & 0.5 & 0.97 & 0.42 &
0.21 & 0.0028 & 0.014 & 0.24 & 2.0 & 0.23 & 0.65\\
{\it m-2} & $m_{lo}=0.43$ & 0.54 & 0.22 & 0.5 & 0.98 & 0.42 &
0.21 & 0.0028 & 0.014 & 0.23 & 2.0 & 0.23 & 0.64\\
{\it m-3} & $m_{br}=1.25$ & 0.53 & 0.22 & 0.5 & 0.98 & 0.42 &
0.21 & 0.0029 & 0.014 & 0.23 & 2.0 & 0.23 & 0.64\\ 
\enddata 
\tablecomments{Canonical and Salpeter models included for comparison
  purposes; $\varepsilon^{Ia}=0.13$ and star/gas=0.25 variations of
  model {\it sl-3} also included.}
\end{deluxetable}

\subsection{Other Variations}

\subsubsection{Supernova Yields}

In most cases, varying the SNIa yield set primarily affects the
predicted (stellar and ICM) Ni/Fe ratios -- which are not
well-determined in clusters at this time. Exceptions are yield sets
with particularly low ($<0.4~{\mathrm M}_{\sun}$) Fe yields, i.e. the
C-DEF and C-DDT in \cite{mae2010}. These models cannot
self-consistently produce the observed stellar abundances and
$Z_{Fe,ICM}=0.5$. Since there is a narrow range of Fe yields in SNcc
calculations, our results are insensitive to the choice of $Z_{cc}$ --
though, in principle, abundance ratios among $\alpha$-elements could
carry SNcc yield diagnostic information. Decreasing the SNcc upper
mass limit, $m_{up}$, with respect to the IMF upper limit, $m_{hi}$,
lowers the predicted metallicities -- though the effect is small
unless the upper mass IMF slope is very flat.

Overall our models are conservative in the sense of maximizing Fe
yields by adopting high values for the SNIa Fe yield ($0.74~{\mathrm
  M}_{\sun}$), and for $m_{hi}$ ($150~{\mathrm M}_{\sun}$).

\subsubsection{SNIa Efficiency}

\cite{maoz2012} estimated that a $1.7\times$ higher Type Ia supernova
rate per unit mass of star formed is implied by the ICM Fe abundances
for otherwise standard assumptions. We construct a scenario along
these lines by considering an increase in $\varepsilon^{Ia}$ -- the
fraction of $3-8~{\mathrm M}_{\sun}$ that explode as SNIa -- from
0.076 to 0.13 (${f_{*}}/{f_{ICM}}$ and stellar abundances constrained
to match the standard values; Section 4.1). Indeed we find
$Z_{Fe,ICM}=0.5$ for an IMF that otherwise (i.e., in terms of IMF,
star formation efficiency, SNcc lock-up fraction and rate, and ICM
primordial fraction) approximates the standard model (Tables 3 and
4). Naturally, the increase in efficiency of formation of SNIa
progenitors results in a large asymmetry between the stellar and ICM
abundance patterns ($\alpha^{SN}=3.1$), as reflected in the more
nearly solar $[\alpha/Fe]_{ICM}$ ratio and consistent with ICM
abundance patterns ($R^{SN}_{ICM}=0.4$; \cite{bsl2012}) -- see Tables
3 and 4, and Figures 9 and 10 that display results for such a
variation of model {\it sl-3}. If this is the correct explanation for
the observed ICM abundances, one must seek an astrophysical
explanation for boosting $\varepsilon^{Ia}$ in rapidly star-forming
systems.

\subsubsection{Star-to-ICM Ratio}

As briefly discussed in Section 2.2, the present-day star-to-ICM ratio
may exceed our standard value of 11\%, e.g. due to an unaccounted-for
ICL fraction or underestimate of the stellar mass-to-light ratio. If
we increase ${f_{*}}/{f_{ICM}}$ from 0.11 to 0.25, a generally
satisfactory resolution of the cluster elemental abundance paradox is
achieved\footnote{Stellar abundances are unchanged.} -- a level of Fe
enrichment and abundance pattern ($R^{SN}_{ICM}=0.32$) consistent with
observations is attained for an IMF and other parameters in line with
expected values -- with the notable exception of the increase in star
efficiency to $\varepsilon_{sf}=32$\% -- see Tables 3 and 4. The
results of this variation on model {\it sl-3} are also displayed
Figures 9 and 10. It is worth pointing out at this juncture that we
defined $\varepsilon_{sf}$ as the fraction of {\it cluster} baryons
that form stars; in our models the fraction of {\it galactic} baryons
that form stars (where ``galaxies'' are defined as locations where
stars form and eject mass into the ICM) is
$\varepsilon_{sf}/\varepsilon_{gal}=(1+\delta_{GW})^{-1}$ -- 75\% for
this ${f_{*}}/{f_{ICM}}=0.25$ model, but also 2/3 for the standard
model.


\begin{figure}[h]
\includegraphics[scale=0.8,angle=0]{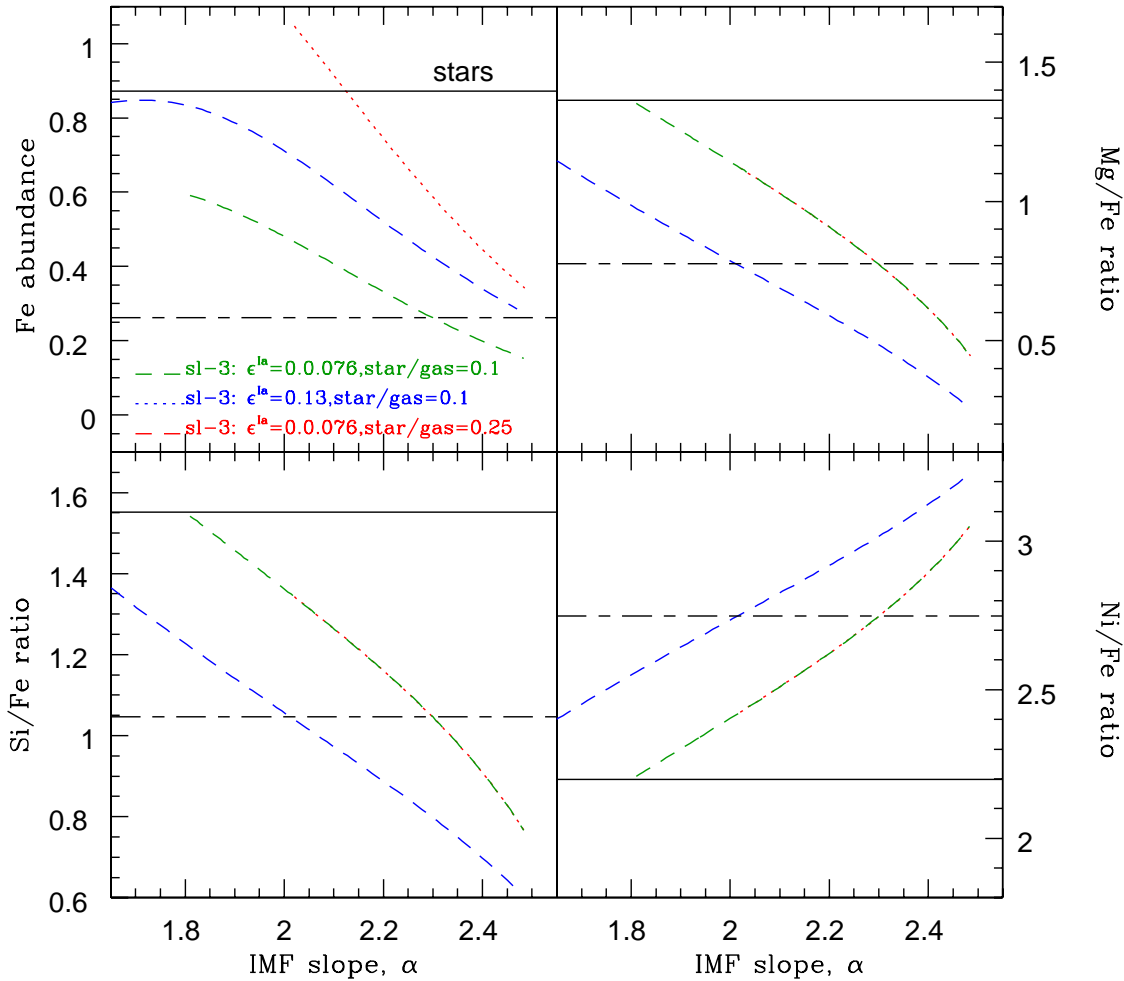} 
\caption{\footnotesize{ICM Fe abundance and Mg/Fe, Ni/Fe, and Si/Fe
    abundance ratios versus slope above $0.5~{\mathrm M}_{\sun}$
    (assuming slope 1.3 for $m<1~{\mathrm M}_{\sun}$) -- i.e., model
    {\it sl-3}, for boosted SNIa progenitor formation efficiency
    ($\varepsilon^{Ia}$) {\it or} star-to-ICM ratio
    (${f_{*}}/{f_{ICM}}$). As for all models, stellar abundances
    (solid horizontal lines) are fixed at their standard values (Table
    1). Results for a standard IMF (broken horizontal lines), and for
    model {\it sl-3} with standard $\varepsilon^{Ia}$ and
    ${f_{*}}/{f_{ICM}}$, are reproduced.}}
\end{figure}

\begin{figure}[h]
\includegraphics[scale=0.25,angle=0]{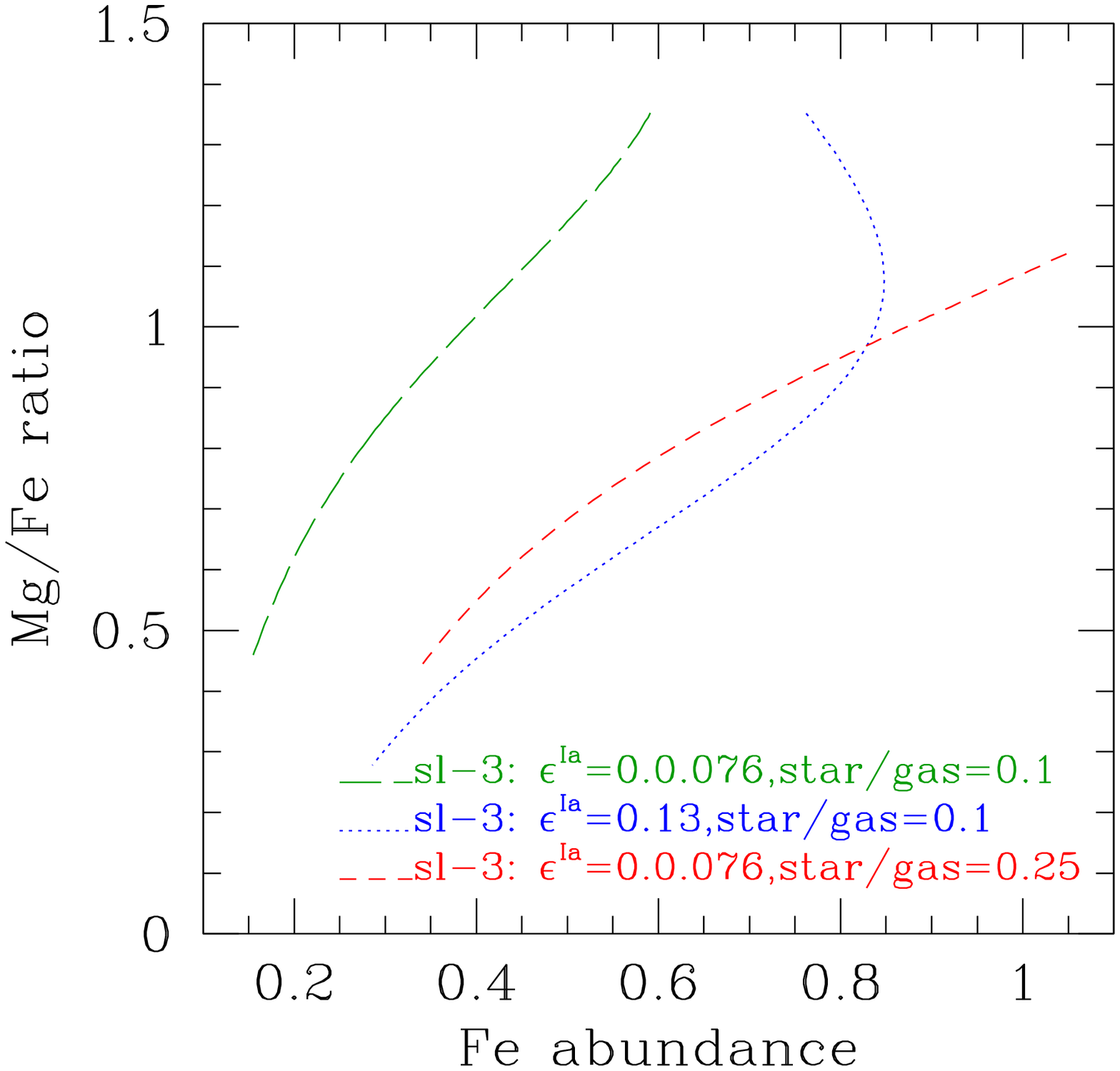}
\hfil
\includegraphics[scale=0.25,angle=0]{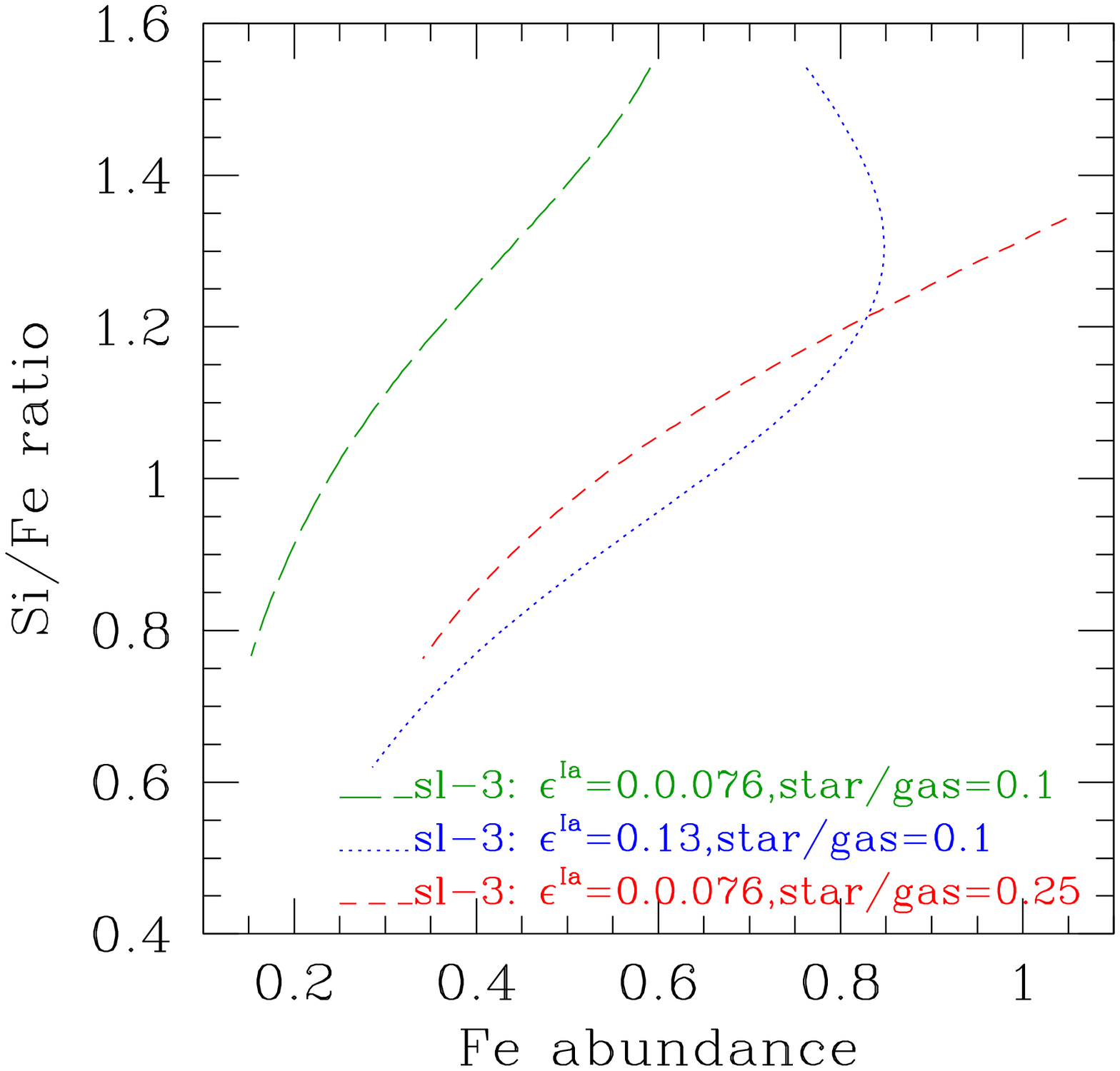}
\hfil
\includegraphics[scale=0.25,angle=0]{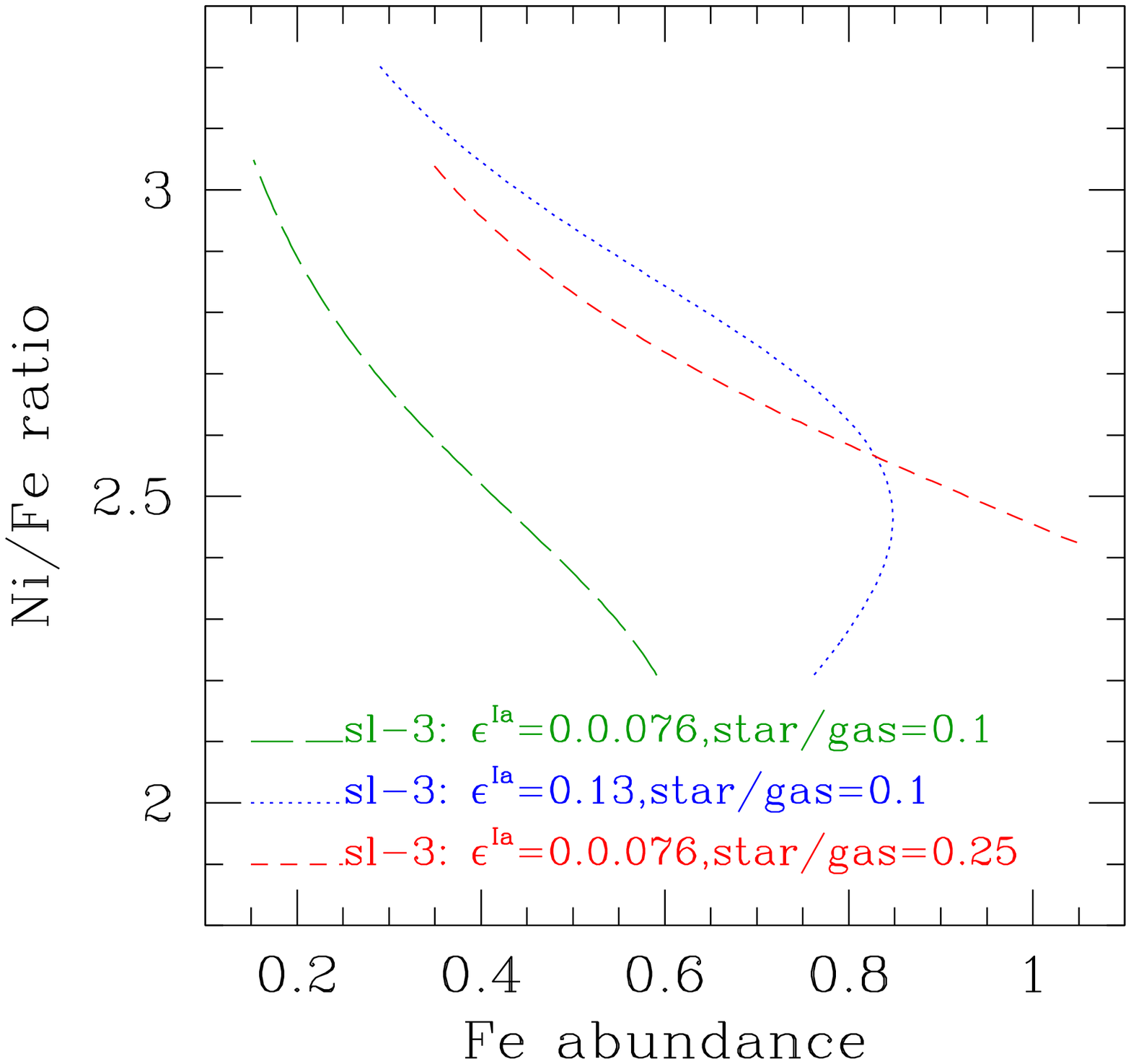}
\hfil
\caption{\footnotesize{ICM abundance ratios with respect to Fe for Mg
    -- {\it left} panel (a), Si -- {\it middle} panel (b), and Ni --
    {\it right} panel (c) -- {\it vs.} Fe abundance for model {\it
      sl-3} and its boosted $\varepsilon^{Ia}$ and star-to-ICM ratio
    counterparts.}}
\end{figure}


\section{\bf Summary and Discussion}

Star formation, with a canonical IMF and standard efficiency in
producing SNIa, that builds up a stellar population comprising $\sim
10$\% of the current overall cluster baryon content falls short by a
factor of $>2$ of explaining a typical rich cluster half-solar ICM Fe
abundance (Sections 2.3-2.4). This is the case even if predicted ICM
abundances are enhanced by increasing the efficiency at which metals
are ejected from galaxies (and where, as a result, the overall
abundance in stars is significantly below solar), unless the
conversion efficiency of cluster baryons into stars is also increased
well above 10\% (Section 3).

Section 4 (and Appendix A) constructs and utilizes a phenomenological
model for the evolution of an old, simple stellar population to
quantify the changes in the IMF shape (high and low mass slopes, break
mass) from its standard form required to bring the ICM metallicity and
cluster stars into concordance in the sense that they be consistent
with the same parent star formation history. The necessary departure
may be either in the ``bottom-light'' or ``top-heavy'' sense, with the
former tentatively preferred based on better agreement with observed
ICM abundance patterns and on a higher primordial ICM fraction. It is
further demonstrated that if a standard IMF is to be preserved, a
boost in the efficiency of forming stars from gas well beyond that
consistent with a gas-to-star ratio of 10, and/or of producing SNIa
progenitor systems, is required. These calculations are conservative
in the sense of maximizing the enrichment of the stellar population
through the choice of SN parameters (e.g., SNIa Fe yields, the upper
mass limit for SNcc).

Stars born in cluster potential wells (or those of their progenitors)
must be responsible for the high level of enrichment measured in the
ICM; however, there is increasing tension between this truism and the
parsimonious assumption that the stars in the generally old
populations studied optically emerged from the same formation sites
during the same epochs. Quantifying this tension, and bolstering the
case against the universality of star formation are the two primary
implications of this study. In the remainder of this section we
elaborate on these themes.

\subsection{The ICM-Enriching Stellar Population as Distinct from that
  Observed in Elliptical Galaxies}

In some cases the departure from the canonical IMF is modest -- a
shift of a few tenths in the slope over some mass range or an increase
in the mass at which the slope steepens from 0.5 to $1.25~{\mathrm
  M}_{\sun}$. However, optical determinations of the IMF from
kinematic and population studies in elliptical galaxies are trending
in the opposite direction
\citep{treu2010,auger2010,vanDokkum2010,vanDokkum2011,vanDokkum2012,conroy2012,thomas2011,dutton2012,smith2012,cappellari2013,tortora2012,spiniello2011,spiniello2012,ferreras2013,sonnenfeld2012,goudfrooij2013,dutton2013,dt2013}. The
kinematic evidence that indicates a larger mass-to-light ratio in
massive ellipticals than expected based on a standard IMF is
consistent with either an IMF that is top-heavy and so produces more
stellar remnants, or one that is bottom heavy and produces more
unevolved low mass stars. However population synthesis modeling of
elliptical galaxy spectra favor the latter, and a consensus appears to
be emerging for an IMF that is bottom-heavy in elliptical galaxies
with central velocity dispersions $>150$ km s$^{-1}$ where most of the
present-day stellar mass in galaxy clusters reside, being at least as
steep as a Salpeter IMF if characterized by a single slope (and
steeper still at the highest galaxy masses). The apparent dependence
of the IMF on elliptical galaxy mass is strong evidence against the
existence of a universal IMF.

The chasm between the amount of metals expected to be produced from a
stellar population with such a steep IMF, and the observed level of
cluster enrichment is illustrated in the plots of Fe abundance in the
ICM versus cluster star-to-gas ratio for three distinct IMFs in Figure
11. Standard yield sets and values of the parameters $\delta_{GW}$
(0.5), $f^{Ia}_p$ (0.5), and $\varepsilon^{Ia}$ (0.076) are assumed
(Section 4.1).\footnote{That is, the solutions are no longer
  constrained to match the standard stellar abundances.} Also plotted
are the overall averaged Fe abundances for the total cluster baryons;
these are independent of the detailed galaxy evolution parameters
$\delta_{GW}$ and $f^{Ia}_p$. As might easily be inferred from
previous considerations, it is clear that the model with Salpeter IMF
requires an excessively large gas-to-star ratio, and that an IMF as
steep as $\phi\sim m^{-3.05}$ unequivocally falls short by more than
an order of magnitude of producing the required amount of metals.

\begin{figure}[h]
\includegraphics[scale=0.8,angle=0]{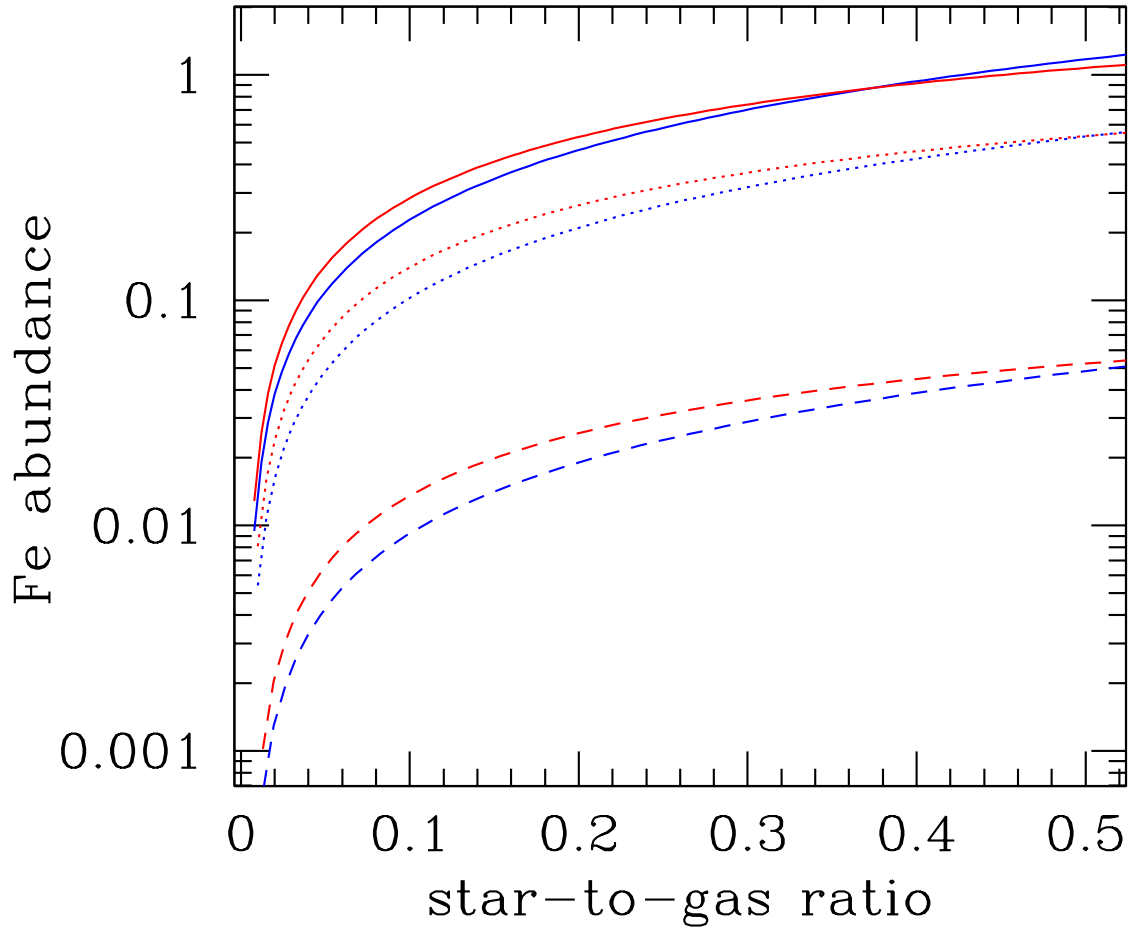}
\caption{\footnotesize{Fe abundance in the ICM (blue curves) and,
    overall, in cluster baryons (red curves) for the following three
    IMFs: (solid curves) standard \citep{kroupa2012}, (dotted curves)
    single slope with $\phi\sim m^{-2.35}$ \citep{salpeter1955},
    (dashed curves) single slope with $\phi\sim m^{-3.05}$. }}
\end{figure}

The most straightforward explanation for reconciling the steep IMFs
based on optical spectroscopic studies of elliptical galaxies and the
relatively flat IMFs needed to produce the cluster metals is to reject
the conventional wisdom that the stellar populations in ellipticals
that dominate the cluster stellar mass are primarily responsible for
ICM enrichment. Such a decoupling begs the question of the origin and
present-day whereabouts of the ICM-enriching stars and motivates
consideration of scenarios with pre-enrichment (that would presumably
be accompanied by pre-heating) in protocluster environments by a
currently inconspicuous stellar population. The independence of the
enriching population of high mass stars from present-day galaxies is
supported by the remarkably small range of cluster metallicities
\citep{baldi2012} as compared with the large variation in inferred
stellar mass fraction -- and the lack of correlation between the two
\citep{bregman2010}.

However, an important caveat with respect to the optical spectroscopic
studies is their general confinement to regions well inside the
half-light radius as well as systems at low redshift. Given the
emerging paradigm of the multi-stage, inside-out
formation/assembly/growth of ellipticals by multiple mechanisms
(star-forming major mergers, ``dry'' minor mergers, and cold and hot
gas accretion; e.g., Conselice et al. 2013, Patel et al. 2013, Huang
et al. 2013), spatial gradients and temporal evolution in properties
of elliptical galaxy stellar populations such as the IMF is to be
expected \citep{labarbera2012}. With the current dearth of global
constraints, as well as degeneracies between the inferred dark matter
content and the IMF \citep{wegner2012,tortora2013} and possible
systematic errors resulting from nonsolar abundance ratios
\citep{ferreras2013}, an IMF in cluster ellipticals that is flatter
than currently inferred in the core when integrated over space and
time is plausible \citep{dave2008,worthey2011,narayanan2012b}.

Recent arguments for a reconsideration of bimodal star formation --
hints for an IMF that is top-heavy in older, but bottom-heavy in
younger, star clusters \citep{zaritsky2012,zaritsky2013}, or top-heavy
in denser environments \citep{marks2012} -- support this. The
evolution in the cluster star formation environment plausibly leads to
an IMF in elliptical galaxies (and their progenitors) that transition
from one initially weighted towards the mass range that includes SNcc
and (prompt) SNIa progenitors to one especially conducive to low mass
star formation at later times (Narayanan \& Dav{\'e}(2012ab; see, also
Suda et al. 2013). We note that the high metallicities seen in the gas
in high redshift quasar hosts \citep{dietrich2003a,dietrich2003b} also
indicate an early epoch of rapid star formation that efficiently
produces SN progenitors and is accompanied by powerful galactic
outflows \citep{dimatteo2004,wang2012}.

\subsection{The Enriching Stellar Population as Distinct from that
  Observed in our Galaxy}

The hypothesis that star formation is universal is refuted by analysis
of the level and pattern of ICM elemental abundances. The star
formation characteristics of the stellar population responsible for
these metals must depart from that studied locally in one or more of
the following ways: (1) engender a higher fraction of high mass stars,
(2) more efficiently form stars from gas, (3) more efficiently produce
SNIa progenitor systems. In addition, we saw in the previous
subsection that the IMF if the enriching population is distinct from
that recently inferred in the central regions of elliptical
galaxies. Arguments for (1) were presented above. We now examine the
feasibility, and implications, of hypotheses (2) and (3).

The true star-to-gas ratio (and implied star formation efficiency)
remains uncertain, with stellar masses difficult to estimate given low
surface brightness extended light and the likelihood of multiple
stellar populations that complicate the conversion from measured light
in some aperture to total stellar mass
\citep{munshi2013,mitchell2013}. The ICL contribution is also
uncertain -- although recent measurements for those most massive
clusters that concern this work indicate modest ICL mass fractions
\citep{kb2007, sanderson2013}. Both bottom-heavy as now being inferred
in elliptical cores, and top-heavy/bottom-light as required by ICM
enrichment, IMFs may result in upward revisions in mass-to-light
ratios. A global value of $f_{*}/f_{ICM}\ge 0.25$, even for rich
clusters, does not seem to be excluded by observations at this
time. The star formation efficiency corresponding to $f_{*}/f_{ICM}\ge
0.25$, excluding the primordial portion of the ICM that does not
engage in star formation (see Section 4.4.3, above), is
$\{5\varepsilon_{gal}(1-r_{*})\}^{-1}$, where $\varepsilon_{gal}$
(first defined in Appendix A) is the fraction of baryons initially in
star-forming structures. This quantity has an absolute minimum of 0.2,
is $>0.3$ for any reasonable value of the mass return fraction
$r_{*}>1/3$, and $>0.5$ for $r_{*}>1/3$ and
$\varepsilon_{gal}<0.6$. These considerations indicate that a
substantial fraction of protocluster gas was in the form of dense
star-forming protogalaxies or pre-galactic fragments. There are
clearly profound implications for such a high stellar fraction and
star formation efficiency for evaluating the magnitude -- and perhaps
even the reality -- of the ``overcooling'' problem
\citep{mccarthy2011} and the physics of star formation quenching as it
pertains to the cluster environment, as well as for the precision in
using cluster gas fractions -- that must be converted to baryon
fractions using a correction for stellar content -- to constrain the
cosmological world model \citep{allen2008}.

Given the uncertainty in the nature and possible diversity of SNIa
progenitors, and the difficulties in reproducing observed rates (e.g.,
Toonen et al. 2012, Quimby et al. 2012, and references therein), the
feasibility of an efficiency of SNIa progenitor formation in galaxy
clusters that exceeds the standard $\varepsilon^{Ia}=0.076$ is not
easily evaluated, but cannot be summarily dismissed. Recent work in
this area provides hints, on the one hand, of a {\it downward}
revision in the global estimate of $\eta^{Ia}$; but, on the other, of
a higher value in galaxy clusters
\citep{perrett2012,mmb2012,graur2013,quimby2012}. Both an IMF that
produces additional stars in the $3-8~{\mathrm M}_{\sun}$ range, and
an increase in $\varepsilon^{Ia}$, may boost the value of $\eta^{Ia}$.

\subsection{Future Directions}

Progress in resolving the cluster elemental abundance paradox will
proceed, in parallel, along theoretical and observational lines as
follows. Since there is data on the spatial distribution of stars and
gas \citep{battaglia2012}, and on the evolution of the Fe abundance
\citep{baldi2012}, we are extending our modeling to multi-zone and
time-dependent treatments -- with particular attention to possible
mechanisms of pre-enrichment and predictions for cluster SN (and
$\gamma$-ray burst rates; see below) as a function of redshift -- that
further constrain enrichment scenarios. We will also extend our
investigation to galaxy groups, including fossil groups.

SN surveys are attaining better statistics, particularly at high
redshift, and are sharpening the accuracy of SN rates, delay-time
distributions, and environmental dependencies. Optical spectroscopic
studies are improving both observationally, and in terms of the
complexity of the stellar population models used to interpret
them. X-ray studies of elliptical galaxy interstellar and
circumstellar gas provide additional probes of elliptical galaxy
evolution \citep{loewenstein2010,loewenstein2012}. Future improvements
in measuring cluster abundance patterns beyond Fe, and in abundance
and abundance pattern gradients and time-variation are crucial.

Finally, we note that many of the mechanisms suggested here for
explaining the level of ICM enrichment -- pre-enrichment by massive
stars, efficient and rapid conversion of stars to gas, an IMF skewed
to high masses -- would suggest that the protocluster environment is a
fertile one for producing $\gamma$-ray bursts
\citep{lloydronning2002,wang2011,elliot2012}, a suggestion we are
following up on.

\subsection{\bf Concluding Remarks}

The goal of this work is to quantify the requirements for the stellar
population responsible for injecting metals into the ICM, and evaluate
the feasibility that the stars we see today originate from the same
source. One is driven to conclude that there is a profound divergence
between the ICM-enriching population and that in the ensemble of
elliptical galaxies based on standard assumptions about, and recent
optical spectroscopic population studies of, the latter. This is
inferred from the number of SN progenitors needed for the former and
that expected in ellipticals based on their integrated light and
apparent bottom heavy IMF, implying the existence of a distinct
``hidden'' stellar source of metals that may or not inhabit the same
space as these galactic stars at the same time. While the modeling
here is basic, the conclusion depend mostly on simple accounting of
metals and unlikely to be altered in more sophisticated
treatments. And although the rich galaxy clusters we consider
represent an extreme environment, there are broader implications for
ellipticals, since mass is a much stronger determinant of their
formation than environment
\citep{gruetzbauch2011a,gruetzbauch2011b}. However in it is in the ICM
where these phenomena are embedded and remain accessible, given the
dominance by elliptical galaxies of cluster light, and the closed-box
nature of these deepest of potential wells.

We present compelling evidence for a diversity of star formation in
terms of some combination of efficiency, IMF, and ability to produce
SNIa progenitors. Implications to be further explored include possible
impacts on using cluster baryon fractions to constrain cosmology,
converting stellar light to mass, and treating star formation and
pre-heating/feedback -- and evaluating overcooling -- in semi-analytic
models of galaxy formation. 

Occam's razor is violated in rich galaxy clusters -- although metals
are made in stars and most of the stars we observe are in elliptical
galaxies, this stellar population as currently understood is evidently
not responsible for producing the metals in the ICM. Moreover, the
nature of the star formation that did produce these metals is clearly
very different from that we are most familar with, as well as that
recently inferred in elliptical galaxies.

\acknowledgments

The author wishes to acknowledge helpful (and enjoyable) discussions
with Richard Mushotzky and Esra Bulbul, and feedback from an anonymous
referee. This paper is dedicated to my late father, Jerry
Loewenstein, for reasons numerous, oblique, and impossible to
articulate.

\appendix

\section{A Simple Model for the Composite Chemical Evolution of Cluster
  Galaxies}

We approximate the stellar population responsible for enriching the
ICM as originating from a single, brief, and early star formation
episode (e.g, Andreon 2013). As discussed in Section 4, we adopt a
three-part, continuous, monotonically decreasing, piece-wise power-law
form for the initial mass function of forming stars (IMF)
$\phi(m)\equiv dN/dm$ following \citet{kroupa2001,kroupa2012}
extending from $m_{lo}$ to $m_{hi}$ and normalized so that
$\int^{m_{hi}}_{m_{lo}}dm\phi(m)=1$. Thus $\phi=kf(x)$ for $x\equiv
m/m_{lo}$, where
\begin{equation}
f(x)= x^{-{\alpha_i}}\prod\limits^{i}_{j=1}c_j, \ \ \ x_i<x\le x_{i+1}, \ \ i=1,3;
\end{equation}
$x_1=c_1=1$, $c_j=c_{j-1}x_j^{{\alpha_j}-{\alpha_{j-1}}}$, $x_4\equiv
m_{hi}/m_{lo}$, and $k$ is determined from the normalization
condition. Thus, in addition to the lower and upper mass limits, three
slopes ($\alpha_1,\alpha_2$, and $\alpha_3$) and two break-masses
($m_2=m_{lo}x_2$ and $m_3=m_{lo}x_3$) must be specified.

For the canonical IMF \citep{kroupa2012} that we adopt as default,
($m_{lo},m_2,m_3,m_{hi}$)= (0.07,0.5,1.0,150), where all masses are in
${\mathrm M}_{\sun}$ and
($\alpha_1,\alpha_2$,$\alpha_3$)=(1.3,2.3,2.3). We consider IMFs where
we vary either $m_{lo}$ (over [0.01,0.5]), $m_{3}$ ([0.5,8]); or
$\alpha_1$, $\alpha_2$, or $\alpha_3$ (all over [0.3,$\infty$)).

The mass return fraction for intermediate mass stars is given by
\begin{equation}
r^{im}_*=\frac{\int^{m_{cc}}_{m_{to}}dm\phi(m)\Delta
    m(m)}{\int^{m_{hi}}_{m_{lo}}dm\phi(m)m},
\end{equation}
where $m_{to}=0.9~{\mathrm M}_{\sun}$ is the main sequence turn-off
mass, $m_{cc}=8~{\mathrm M}_{\sun}$ is the lower mass limit for SNcc
progenitors that we use to delineate ``intermediate'' and ``high''
mass stars, and
\begin{equation}
\Delta m(m)=m-m_{rem,wd}(m),
\end{equation}
where 
\begin{equation}
m_{rem,wd}=0.394+0.199m
\end{equation}
is the white dwarf remnant mass \citep{kalirai2008}.
Similarly, for high mass stars 
\begin{equation}
r^{hi}_*=\frac{\int^{m_{up}}_{m_{cc}}dm\phi(m)\Delta
    m(m)}{\int^{m_{hi}}_{m_{lo}}dm\phi(m)m},
\end{equation}
where $\Delta m(m)$ is now based on an averaged remnant mass (see
below),
\begin{equation}
\Delta m(m)=m-m_{rem,SNcc}(m).
\end{equation}
 The specific number of SNcc explosions per star formed is
\begin{equation}
\eta^{cc}=\frac{\int^{m_{up}}_{m_{cc}}dm\phi(m)}{\int^{m_{hi}}_{m_{lo}}dm\phi(m)m}.
\end{equation}

A simple chemical evolution model for the composite stellar population
in clusters is constructed and adopted. We use this to calculate the
total mass return from high mass stars as the metallicity of the
progenitor stellar population is built up, and to connect the ICM
enrichment parameters with the astrophysics of the formation of
cluster galaxies. The model is appropriate for stellar populations
where conversion of gas to stars is relatively rapid and efficient,
and so may be applied to cluster galaxies where ellipticals dominate
the stellar mass and star formation is accelerated in general due to
the high primordial overdensity. As such, galaxy evolution is divided
into two epochs: active and passive, and three phases
\citep{loewenstein2006}: star-forming gas (``ISM''), stars, and
non-star-forming gas (``ICM''). Note that any hot halo gas --
relatively insignificant in mass compared to the ``true'' ICM for rich
clusters -- is subsumed under the ICM category. In the active phase,
all star formation and SNcc explosions occur and all the initial (ISM)
mass in galaxies is consumed by star formation or ejected by galactic
winds. In the passive phase, stellar mass return and delayed SNIa
continue to enrich the ICM.

The cluster as a whole is treated as a closed box, with mass and metal
exchange among the phases and metal production by the stellar
component. We model the active phase essentially following the
prescription of \cite{qian2012} for the case of no infall. Mass
return is neglected in their approach, and we correct for this in the
passive phase. Respectively, the evolution equations for the mass in
stars, ISM, and ICM are as follows:
\begin{equation}
\frac{dM_*}{dt}=\dot{M}_{SF},
\end{equation}
\begin{equation}
\frac{dM_{ISM}}{dt}=-\dot{M}_{SF}-\dot{M}_{GW},
\end{equation}
\begin{equation}
\frac{dM_{ICM}}{dt}=\dot{M}_{GW}.
\end{equation}
It is assumed that the rate of outflow is proportional to the rate of
star formation that is, in turn, proportional to the ISM mass:
$\dot{M}_{\rm SF}=\lambda_{SF}M_{ISM}$, $\dot{M}_{\rm
  GW}=\delta_{GW}\dot{M}_{SF}=\lambda_{SF}\delta_{GW}M_{ISM}$, so that
the solutions to equations (A8)-(A10) are
\begin{equation}
M_{ISM}=M_{GAL0}e^{-\lambda t},
\end{equation}
\begin{equation}
M_*=M_{GAL0}(1+\delta_{GW})^{-1}(1-e^{-\lambda t}),
\end{equation}
and
\begin{equation}
M_{ICM}=M_{ICM0}+M_{GAL0}\delta_{GW}(1+\delta_{GW})^{-1}(1-e^{-\lambda t}),
\end{equation}
where $\lambda\equiv \lambda_{SF}(1+\delta_{GW})$, and initial
conditions correspond to masses of $M_{GAL0}$ in star forming gas
(presumably, mostly in galaxies) and $M_{\rm ICM0}$ in the ICM, and
$M_*=0$.

The evolution equations of the corresponding ``ith'' element mass
fractions are as follows:
\begin{equation}
\frac{df_{i,ISM}}{dt}=\frac{{\dot{M}}_{i,SN}}{M_{ISM}},
\end{equation}
\begin{equation}
\frac{df_{i,*}}{dt}=\frac{{\dot{M}}_{SF}}{M_*}
(f_{i,ISM}-f_{i,*}),
\end{equation}
and
\begin{equation}
\frac{df_{i,ICM}}{dt}=\frac{{\dot{M}}_{GW}}{M_{ICM}}
(f_{i,ISM}-f_{i,ICM}).
\end{equation}
${\dot{M}}_{i,SN}$ (the only source term in the set of equations) is
the nucleosynthetic production of the ith element,\footnote{Only
  elements primarily synthesized by massive stars and SNIa are
  considered here.}
\begin{equation}
{\dot{M}}_{i,SN}=\delta_{i,SN}{\dot{M}}_{SF},
\end{equation}
\begin{equation}
\delta_{i,SN}\equiv \eta^{cc}\langle y_i^{cc}\rangle+f^{Ia}_p\eta^{Ia}y_i^{Ia},
\end{equation}
where, as previously defined (see Section 2), $y_i^{Ia}$ and $\langle
y_i^{cc}\rangle$ are the yields per SNIa and SNcc, $\eta^{cc}$ and
$\eta^{Ia}$ the numbers of SNcc and SNIa explosions per star formed;
and, $f^{Ia}_p$ is the SNIa fraction considered ``prompt'' in the
sense that they occur during the star formation epoch (not necessarily
part of a distinct prompt SNIa mode).

Analytic solutions for the metal mass fractions are as follows:
\begin{equation}
f_{i,ISM}=\delta_{i,SN}\lambda_*t
\end{equation}
\begin{equation}
f_{i,*}=\delta_{i,SN}(1+\delta_{GW})^{-1}\frac{1-e^{-\lambda
    t}(1+\lambda t)}{1-e^{-\lambda t}},
\end{equation}
and
\begin{equation}
f_{i,ICM}=\delta_{i,SN}\delta_{GW}(1+\delta_{GW})^{-2}\frac{M_{\rm
      GAL0}}{M_{ICM}(t)}\left[{1-e^{-\lambda t}(1+\lambda t)}\right],
\end{equation}
assuming negligible pre-enrichment of any phase.

Masses and metallicities at the end of the active phase are assigned
according to the $t\rightarrow\infty$ limit of equations
(A10)-(A12),(A19)-(A21), following the presumption that most star
formation in clusters occurs on a timescale much shorter than the
current cluster age, and then adjusted for the ensuing passive
injection of stellar mass return and ``delayed'' SNIa. The final
stellar mass and abundances (including remnants) are, therefore, given
by
\begin{equation}
M_*=M_{GAL0}(1-r_{*})(1+\delta_{GW})^{-1},
\ \ \ f_{i,*}=(1+\delta_{GW})^{-1}(\eta^{cc}\langle
y_i^{cc}\rangle+f^{Ia}_p\eta^{Ia}y_i^{Ia}),
\end{equation}
and the final ICM mass and abundances (elemental mass fractions) by
\begin{equation}
M_{ICM}=M_{ICM0}+M_{GAL0}(1+\delta_{GW})^{-1}(\delta_{GW}+r_{*})
\end{equation}
and
\begin{eqnarray}
f_{i,ICM}&=&f_0(1+\delta_{GW})^{-1}[1+\delta_{GW}+f_0(\delta_{GW}+r_{*})]^{-1}
\{(\delta_{GW}+r_{*})\eta^{cc}\langle
y_i^{cc}\rangle\nonumber \\ 
&+&[(1+\delta_{GW})-(1-r_{*})f^{Ia}_p]\eta^{Ia}y_i^{Ia}\}
\end{eqnarray}
where the total mass return fraction is $r_*=r^{hi}_*+r^{im}_*$. For
the overall baryon metallicity
\begin{equation}
f_{i,bar}=\varepsilon_{sf}(\eta^{cc}\langle y_i^{cc}\rangle+\eta^{Ia}y_i^{Ia}),
\end{equation}
where the star formation efficiency defined in equation (1) is related
to $\varepsilon_{gal}$, the ``galaxy formation efficiency'' (the
fraction of baryons initially in star-forming structures) according to
\begin{equation}
\varepsilon_{sf}=\varepsilon_{gal}(1+\delta_{GW})^{-1},
\end{equation}
where $\varepsilon_{gal}=f_0/(1+f_0)$ and $f_0=M_{\rm GAL0}/M_{ICM0}$.

This formalism enables us to interpret the supernova lock-up
parameters introduced in Section 2.4 in the context of the chemical
evolution of clusters galaxies and place them on a firmer physical
footing. Naturally the supernova asymmetry parameter
$\alpha^{SN}={(f^{Ia}_p)}^{-1}$; i.e., it is the inverse of the
fraction of SNIa that explode during the epoch when cluster stars
form. Our Sections 3 and 4 default $\alpha^{SN}=2$ is consistent with
estimates of the prompt SNIa fraction (Maoz et al. 2011, and
references therein; Grauer \& Maoz 2013). The lock-up fraction may be
expressed as $\beta^{cc}=(1-r_{*})(1+\delta_{GW})^{-1}$, and naturally
depends both on how efficiently stars lose mass, and how efficiently
star formation induces galactic winds.

\subsection{Mass Return for Massive Stars}

The active phase chemical evolution model is utilized to calculate the
total mass return from high mass stars -- which can be substantial for
top-heavy IMFs. This is motivated by the profound impact of
metallicity on mass loss in massive stars and, hence, fallback and
final remnant mass
\citep{woosley2002,nomoto2006,zhang2008,fryer2012}. The distribution
of forming stars as a function of time and mass \citep{qian2012} is
\begin{equation}
\frac{d^2N}{dmdt}=\phi(m)\frac{{\dot{M}}_{SF}}{\int^{m_{hi}}_{m_{lo}}dm\phi(m)m},
\end{equation}
from which it follows from the expression for the metallicity of star
forming gas, equation (A18), that
\begin{equation}
\frac{d^2N}{dmdf_{Fe}}=\frac{d^2N}{dmdt}\left(\frac{df_{Fe}}{dt}\right)^{-1}=\frac{N_{tot}}{f_{Fe0}}e^{-f_{Fe}/f_{Fe0}}\phi(m),
\end{equation}
where $f_{Fe0}\equiv \delta_{Fe,SN}/(1+\delta_{GW})$. This distribution
is used to calculate the mass return from massive stars through
numerically computing the average remnant mass:
\begin{equation}
m_{rem,SNcc}=\frac{\int^{1}_{0}df_{Fe}\int^{m_{up}}_{m_{cc}}dm\frac{d^2N}{dmdf_{Fe}}m_{rem,SNcc}(m,f_{Fe})}{\int^{1}_{0}df_{Fe}\int^{m_{up}}_{m_{cc}}dm\frac{d^2N}{dmdf_{Fe}}}
\end{equation}
where the remnant mass as a function of mass and metallicity,
$m_{rem,SNcc}(m,f_{Fe})$, is adapted from \citep{fryer2012} (delayed
explosion scenario) using Fe as a proxy for metallicity.

\subsection{IMF-dependence of SNIa Rate}

The number of SNIa explosions per star is expected to vary with IMF as
follows:
\begin{equation}
\eta^{Ia}=\varepsilon^{Ia}\frac{{\int^{{m^{Ia}_{up}}}_{{m^{Ia}_{lo}}}}dm\phi(m)}{\int^{m_{up}}_{m_{lo}}dm\phi(m)m},
\end{equation}
where $m^{Ia}_{up}=8~{\mathrm M}_{\sun}$, $m^{Ia}_{lo}=3~{\mathrm
  M}_{\sun}$, and $\varepsilon^{Ia}=0.076$ yields the observationally
estimated fraction of 3-8$~{\mathrm M}_{\sun}$ stars that explode as
SNIa \citep{maoz2012}.

\end{document}